\documentclass[journal=jacsat,manuscript=article]{achemso}

\usepackage[version=3]{mhchem} 
\usepackage{xcolor}
\usepackage{graphicx}

\author{Yu-Tsung Lin}
\affiliation{Department of Physics, National Cheng Kung University, 701 Tainan, Taiwan.}

\author{Shih-Yang Lin}
\affiliation{Department of Physics, National Cheng Kung University, 701 Tainan, Taiwan.}
\email{sylin.1985@gmail.com}

\author{Yu-Huang Chiu}
\affiliation{Department of Applied Physics, National Pingtung University, 900 Pingtung , Taiwan}

\author{Ming-Fa Lin}
\affiliation{Department of Physics, National Cheng Kung University, 701 Tainan, Taiwan.}
\email{mflin@mail.ncku.edu.tw}

\title[An \textsf{achemso} demo]
  {Alkali-induced rich properties in graphene nanoribbons: Chemical bonding
}

\abbreviations{GNR}

\begin{document}





\begin{abstract}
The alkali-adsorbed graphene nanoribbons exhibit the feature-rich electronic and magnetic properties. From the first-principles calculations, there are only few adatom-dominated conduction bands, and the other conduction and valence bands are caused by carbon atoms. A lot of free electrons are revealed in the occupied alkali- and carbon-dependent conduction bands. Energy bands are sensitive to the concentration, distribution and kind of adatom and the edge structure, while the total linear free carrier density only relies on the first one. These mainly arise from a single $s-2p_z$ orbital hybridization in the adatom-carbon bond. Specifically, zigzag systems can present the anti-ferromagnetic ordering across two edges, ferromagnetic ordering along one edge and non-magnetism, being reflected in the edge-localized energy bands with or without spin splitting. The diverse energy dispersions contribute many special peaks in density of states. The critical chemical bonding and the distinct spin configuration could be verified from the experimental measurements.

\end{abstract}

\section{Introduction}
The graphene-based systems, which are formed by the planar $sp^2$ bondings of carbon atoms, include layered graphites,\cite{tuinstra1970raman,huang1997collective} few- and multi-layer graphenes,\cite{novoselov2004electric,novoselov2005two} one-dimensional graphene nanoribbons (1D GNRs),\cite{son2006half,li2008chemically} and 1D carbon nanotubes (CNTs).\cite{iijima1991helical,iijima1993single} From a geometric point of view, each GNR could be regarded as a finite-width graphene strip or an unzipped CNT. The 1D GNRs have stirred a lot of studies, mainly owing to the complex relations among the honeycomb lattice, the one-atom thickness, the finite-size quantum confinement, and the edge structure. They could be successfully synthesized by the various experimental techniques. The most common methods are to cut graphene, achieved by a metal-catalyzed cutting,\cite{datta2008crystallographic,ci2008controlled} oxidation cutting,\cite{mcallister2007single,fujii2010cutting} lithographic patterning and etching,\cite{han2007energy,chen2007graphene} sonochemical breaking,\cite{li2008chemically,wang2008room} and molecular precursors.\cite{ruffieux2012electronic,huang2012spatially} The available routes from the unzipping of multi-wall CNTs cover strong chemical reaction,\cite{kosynkin2009longitudinal,cataldo2010graphene} laser irradiation,\cite{kumar2011laser} metal-catalyzed cutting,\cite{elias2009longitudinal,parashar2011single} plasma etching,\cite{jiao2009narrow,jiao2010aligned} hydrogen treatment and annealing,\cite{talyzin2011hydrogenation} unzipping functionalized CNTs by scanning tunneling microscope (STM) tips,\cite{paiva2010unzipping} electrical unwrapping by transmission electron microscopy (TEM),\cite{kim2010graphene} intercalation and exfoliation,\cite{cano2009ex} and electrochemical unzipping.\cite{shinde2011electrochemical} The other techniques involve chemical vapor deposition\cite{campos2008bulk} and chemical synthesis.\cite{yang2008two} GNRs exhibit the feature-rich essential properties, such as, electronic structures,\cite{son2006half,chung2016electronic} magnetic properties,\cite{kim2008prediction,chung2016electronic} optical spectra,\cite{yang2007excitonic,lin2000optical} and transport properties.\cite{li2008role,basu2008effect} The electronic properties are diversified by changing by the ribbon width (W),\cite{son2006energy,barone2006electronic} edge structure,\cite{son2006energy,ritter2009influence} edge-passivated dopants,\cite{chang2014geometric,lin2015adatom} adatom adsorptions,\cite{johnson2010hydrogen,lin2013graphene} layer numbers,\cite{huang2008magnetoabsorption} stacking configurations,\cite{sadeghi2012channel} surface curvatures,\cite{lin2012curvature,chang2012curvature} mechanical strains,\cite{chang2007deformation,lin2015feature} electric fields,\cite{chang2006electronic,kan2007will,raza2008armchair} and magnetic fields.\cite{huang2007magnetic,liu2008strong,chung2016electronic} GNRs are expected to be more potentially applicable in future nanodevices.\cite{obradovic2006analysis,wang2008room,saffarzadeh2011spin} In this work, the first-principles calculations are used to investigate the adatom-enriched electronic properties of the alkali-adsorbed GNRs. Whether the alkali adatoms can create the high free electron density will be explored in detail.

GNRs, with or without hydrogen passivation at boundaries, exhibit the semiconducting behavior, as indicated from the theoretical predictions\cite{son2006energy,barone2006electronic,han2007energy} and the experimental measurements.\cite{tapaszto2008tailoring,huang2012spatially,sode2015electronic,chen2013tuning} There are two typical types of achiral GNRs, namely, armchair and zigzag GNRs. The former have the diverse energy gaps ($E_gs$) inversely proportional to widths.\cite{son2006energy,barone2006electronic} Specifically, $E_g$ in the latter, which arises from the occupied and the unoccupied edge-localized energy bands, is induced by the anti-ferromagnetic spin configuration across the nanoribbon.\cite{son2006half,son2006energy} The strong dependence of $E_g$ on W has been confirmed by the electrical conductance\cite{han2007energy,li2008chemically} and tunneling current\cite{tapaszto2008tailoring,huang2012spatially,sode2015electronic,chen2013tuning} measurements. The semiconductor-metal transition is revealed under a transverse electric field.\cite{son2006half} The similar phenomena could also be observed in some edge-decorated GNRs, e.g., Li-, Be-, Na- and K-decorated unzipping GNRs.\cite{lin2015adatom,mao2013edge,chang2014geometric}

There are some theoretical\cite{wang2013adsorption,da2014optical,uthaisar2009lithium} and experimental\cite{johnson2010hydrogen,lin2013graphene} studies on the atom/molecule adsorptions on the planar GRNs. The first-principles calculations on the Co/Ni-adsorbed GNRs show the spin-split energy bands with the metallic behavior.\cite{wang2013adsorption} The adsorption of ligand-protected aluminum clusters on armchair GNRs is predicted to exhibit the semiconducting or metallic band structures, depending on their kinds.\cite{da2014optical} The lithium-adsorbed zigzag GNRs can induce the position-dependent spin configurations.\cite{uthaisar2009lithium} On the experimental side, hydrogen molecules are successfully adsorbed on the Pd-functionalized multi-layer GNRs,\cite{johnson2010hydrogen} and tin oxide nanoparticles are synthesized on GRNs to form a composite material.\cite{lin2013graphene} The former and the latter could be utilized as chemical gas sensors and anode materials in lithium ion battery, respectively. However, a systematic theoretical study on the alkali-adsorbed GNRs is absent up to now. The critical orbital hybridization in the alkali-carbon bonds responsible for the creation of conduction electrons, the relations between the free carrier density and the kind, distribution and concentration of adatom, and the dependence of electronic and magnetic properties on the edge structure, width and adatom position are worthy of a detailed examination.

In this paper, the geometric, electronic and magnetic properties of alkali-adsorbed GNRs with various widths are investigated in detail by the first-principles calculations. The ribbon widths, the armchair and zigzag edges, and the kinds, optimal positions, relative distances, single- and double-side adsorptions and concentrations of alkali adatoms, are included in the detailed calculations. There exist the cooperative or competitive relations among the width-dependent quantum confinement, the spin configurations, and the orbital hybridizations in the alkali-carbon bonds. The critical chemical bondings could be analyzed from the atom-dominated energy bands, the spatial charge distribution, and the orbital-projected density of states (DOS). This work shows that they are responsible for the feature-rich electronic properties. All the conduction and valence bands mainly originate from carbon atoms except for few alkali-dominated conduction bands. The high free electron density in the occupied carbon- and alkali-related conduction bands only depends on the adatom concentration. Specifically, the zigzag systems can exhibit three kinds of spin configurations with the distinct edge-localized energy bands, depending on the alkali positions. Moreover, the 1D metallic behavior are clearly evidenced in a high DOS at $E_F$, accompanied with several prominent peaks. The predicted main features in energy bands and DOS could be examined by angle-resolved photoemission spectroscopy (ARPES) and scanning tunneling spectroscopy (STS), respectively. These rich fundamental features in alkali-adsorbed GNRs are expected to provide potential materials applications in electronic\cite{obradovic2006analysis,wang2008room} and spintronic\cite{saffarzadeh2011spin} devices.

\section{Computational Methods}

The first-principles calculations on the geometric and electronic properties of graphene nanoribbons are performed by using Vienna \emph{ab initio} simulation package.\cite{kresse1996efficient} The exchange-correlation energy of interacting electrons is evaluated from the Perdew-Burke-Ernzerhof functional\cite{perdew1996generalized} under the generalized gradient approximation. Furthermore, the electron-ion interactions are characterized by the projector-augmented wave pseudopotentials.\cite{blochl1994projector} In the calculations of wave functions, plane waves have an maximum energy cutoff of $500$ eV. The vacuum distances between two neighboring nanoribbons along the z-axis and y-axis are, respectively, set to be at least $15$ {\AA} and $12$ {\AA}.
The first Brillouin zone is sampled in a Gamma scheme along the periodic direction by $21\times1\times1$ k points for all structure relaxations, and by $200\times 1\times1$ for further studies on electronic properties. Setting the Hellmann-Feynman forces weaker than $0.01$ eV/{\AA} and the total energy difference of $\Delta\,E_t<10^{-5}$ eV are to determine the convergence criterion for structure relaxation. Regarding Gaussian smearing, the width of $0.05$ eV is taken for the density of states (DOS).

\section{Results and discussion}

The essential properties of the alkali-adsorbed graphene nanoribbons are investigated for different adatoms, concentrations, distributions, widths, and edge structures. For armchair and zigzag GNRs, their widths are, respectively, characterized by the number of dimer lines and zigzag lines along $\hat y$ ($N_A$ and $N_Z$). The former are chosen for a model study in the alkali-induced free carriers. The optimal adatom position, as shown in Fig. 1(a), is situated at the hollow site, almost independent of the above-mentioned five critical factors. The similar position in 2D graphene has been verified by low-energy electron microscopy.\cite{virojanadara2010epitaxial} The adatom height strongly depends on the atomic number, in which $h_{ad}$ varies from 1.75 \AA{} to 3.06 \AA{} as Li $\rightarrow$ Cs (Table 1). Also, the alkali-carbon (A-C) bond lengths grow with the radius of adatom, ranging from 2.23 \AA{} to 3.37 \AA{}. These clearly illustrate that Li adatoms have the strongest bondings with carbon atoms among the alkali-adsorbed systems. GNRs remain the planar structure so that the $\sigma$ bonding due to ($2s, 2p_x, 2p_y$) orbitals of carbon atoms is almost unchanged after alkali adsorption. However, the C-C bond lengths are sensitive to the positions of carbon atoms, but not those of alkali adatoms. In addition, the total ground state energy only presents a negligible change (about several meVs) during the variation of adatom position.

Electronic structures of GNRs are dramatically changed by the alkali-atom adsorption, especially for free electrons in conduction bands. Pristine armchair nanoribbons exhibit a lot of 1D energy bands, as shown in Fig. 2(a) for the $N_A=12$ armchair GNR. Most of them belong to parabolic bands, while few of them have partially flat dispersions within a certain range of $k_x$ (e.g., $E^v=-2.1$ eV, $-4.7$ eV; $-5.7$ eV). All the energy dispersions present the monotonous $k_x$-dependence except for the subband anti-crossings. The occupied valence bands are asymmetric to the unoccupied conduction bands about the Fermi level ($E_F=0$). There is a direct energy gap of $E_g=0.61 $ eV at the $\Gamma$ point, mainly owing to the effect of quantum confinement. The electronic states, with $|E^{c,v}|\le\, 2$ eV, come from the $\pi$ bondings of the parallel $2p_z$ orbitals, and the others are closely related to the $\sigma$ bondings of ($2s$, $2p_x$, $2p_y$) orbitals. In general, the band-edge states occur at $k_x=0$ and 1 (in unit of $\pi\,/3b$). Such critical points are also revealed at other $k_x$'s in the presence of subband anti-crossings.

The alkali-adsorbed graphene nanoribbons exhibit the similar band structures, as clearly indicated in Figs. 2(b)-2(f). The conduction bands are easily modulated by the alkali adatoms. The energy dispersions of valence states are changed by the adatom adsorption, while their number and the dominance of carbon atoms keep the same. These are mainly determined by the orbital hybridizations in A-C bonds bonds (details in Fig. 5). The Fermi level ($E_F$) is changed from the middle of energy gap into the conduction bands, directly reflecting the high charge transfer from alkali atoms to adsorbed systems. The Fermi momentum can characterize the linear free carrier density arising from each occupied conduction band by the relation $\lambda=2k_F/\pi$. There are some extra conduction bands which are created by the alkali adatom. Specifically, there are two or three conduction bands intersecting with the $E_F$, depending the kinds of alkali adatoms. The lowest and/or highest ones are dominated by carbon atoms, and another one by adatoms (dominance proportional to the radius of green circle). The adatom-induced conduction bands are quite different among the various alkali systems, but the distinct systems possess the almost same free electron density (discussed in Table I). In addition, energy bands almost keep the same during the variation of the single-adatom position (not shown).

The electronic structures are enriched by the concentration, relative position, single- and double-side adsorption, and edge structure. When two Li adatoms are situated at the distinct edges of the same side, there are five conduction bands with free electrons, as shown in Fig. 3(a). They only make important contributions for two low-lying conduction bands. The relative position can drastically alter conduction bands near $E_F$. For two neighboring adatoms ((3,7)$_{single}$ in Fig. 1), only one Li-dominated conduction band has free carriers, while another higher-energy one is fully unoccupied (Fig. 3(b)). It should be noticed that the total sum of the Fermi momenta is hardly affected by the relative position. This clearly indicates that the interactions between two alkali adatoms are not responsible for the free carrier density. As to the variation from the single- to double-side absorptions, the conduction bands present an obvious change except for the two-adatom case with the sufficiently long distance. For example, bands structures are different from each other in the single- and double-side adsorptions with two adatoms close to the ribbon middle (Figs. 3(b) and 3(d)), and the opposite is true for the long-distance case (Figs. 3(a) and 3(c)). With the increasing concentration, there are more free carriers in the alkali-created and carbon-dependent conduction bands, as revealed in Figs. 3(e) and 3(f). In addition, the adatom dominance on the valence bands gradually grows, mainly owing to more C-alkali interactions (the increased radii of green circles).

The edge structure can dramatically change the electronic structures in the absence and presence of adatom adsorption. Zigzag GNRs are in sharp contrast with armchair GNRs. The former possess the anti-ferromagnetic ordering across the nanoribbon and the ferromagnetic configuration at each zigzag edge, as clearly illustrated in Figs. 4(e). Zigzag GNRs present a pair of partially flat energy bands nearest to $E_F$ (blue triangles in Fig. 4(a)), in which their wavefunctions are localized at edge boundaries.\cite{son2006energy} That is, such energy bands mainly come from the edge carbon atoms. Their energy dispersions become strong at large $k_x$'s, in which the extra band-edge states come to exist there. The similar states are also revealed in alkali-adsorbed zigzag GNRs (Figs. 4(b) and 4(c)). Both partially flat energy dispersions and band-edge states are expected to induce the distinct special structures in DOS. Furthermore, they can create a direct energy gap at $k_x=0.5$ in the anti-ferromagnetic configuration, e.g., $E_g=0.46$ eV for a $N_Z=8$ ZGNR. It should be noticed that energy gap is determined by the strong competition between spin configuration and quantum confinement. The energy bands are doubly degenerate in the spin degree of freedom, independent of the spin-up- and spin-down-dominated configurations.

Band structures and spin configurations are very sensitive to the changes in the distribution and concentration of adatom. When one alkali adatom is located near the zigzag edge (Fig. 4(b)), the spin-degenerate electronic states become split. The two edge-localized energy bands, which intersect with the Fermi level, have an obvious splitting of $\sim0.7$ eV. However, another two ones below $E_F$ only exhibit a weak splitting. The former and the latter are, respectively, dominated by edge carbon atoms away from and near adatoms. These clearly indicate that the spin configurations are strongly suppressed by the A-C interactions, as revealed from Fig. 4(f). There exists the ferromagnetic ordering at one edge without adatoms. This is responsible for the spin-split energy bands which present more free carriers in the spin-down conduction band (the red curve crossing $E_F$). On the other hand, the spin degeneracy and the anti-ferromagnetic configuration are recovered, while the adatoms are situated at the ribbon center. The absence of spin distributions near adatoms, as shown in Fig. 4(g), further illustrates the magnetic suppression. According to the above-mentioned specific relation between spin distribution and adatom position, the spin-dependent properties are identified to be disappeared for two adatoms at distinct edges, e.g., the spin-degenerate energy bands without magnetism in Fig. 4(d). Furthermore, the two edge-localized energy bands are merged together at $E^v=-1.35$ eV for $|k_x|\le\,0.5$. In addition, the typical magnetic momenta of the edge carbon atoms are about 0.10 $\sim$ 0.15 $\mu_B$ under the ferromagnetic ordering along the zigzag edge. In short, three kinds of magnetic configurations in alkali-adsorbed zigzag GNRs, the anti-ferromagnetic ordering across the ribbon, ferromagnetic ordering only along one edge and non-magnetism, are mainly determined by the adatom positions. Specifically, the second kind of spin configuration corresponds to the spin-split energy bands with the different free electron densities. Such electrons under transport measurements are expected to create the spin-polarized currents, indicating potential applications in spintronic devices.\cite{saffarzadeh2011spin}

The adatom-induced linear free electron density deserves a closer examination. By the detailed analyses and calculations, the total carrier density in conduction bands below $E_F$ is just equal to the adatom density in a unit cell. This is independent of the kinds of adatoms, the adatom positions, the single- or double-side adsorptions, the edge structures, and the ribbon widths, as clearly revealed in Tables 1 and 2. These indicate that the orbital hybridizations in alkali-C bonds (or the $\pi$ bondings in Fig. 5) are almost the same under various adatom adsorptions. The alkali-created free electron density is estimated to be about $\lambda\,\sim 2.31-2.37\times\,10^7$ e/cm for a single adatom in an armchair unit cell, and it can reach $\lambda\,\sim 1.40\times\,10^8$ e/cm for the double-side adsorption of six adatoms in $N_A=12$ armchair GNR. Specifically, the alkali adatoms are deduced to contribute the outmost s-orbital electrons as free carriers in adsorbed systems. The similar results could be generalized to alkali-adsorbed 2D graphenes, i.e., the 2D electron density is dominated by the adatom density, but not the distribution configurations. On the other hand, the charge transfer between alkali and carbon atoms could also be obtained from the Bader analysis. However, it is sensitive to the changes in the kind, position and concentration of adatom, directly reflecting that this analysis cannot be utilized to evaluate the free carrier density in adatom-adsorbed GNRs. The free electron density could be greatly enhanced by the increase of adatom concentration, so that the electrical conductance is expected to behave so. The alkali-doped GNRs might be promising materials in nanoelectronic devices\cite{obradovic2006analysis,wang2008room} or next-generation supercapacitors, e.g., the ultrafast rechargeable metal-ion battery,\cite{lin2015ultrafast,rani2013fluorinated} and the large reversible lithium storages.\cite{paek2008enhanced,wang2009graphene}

Recently, ARPES has emerged as the most powerful experimental technique in the identification of the wave-vector-dependent electronic structures. The experimental measurements on the graphene-related systems could be used to explore the feature-rich band structures under the different dimensions,\cite{sugawara2006fermi,gruneis2008electron,ohta2007interlayer,siegel2013charge,bostwick2007quasiparticle,ohta2006controlling,coletti2013revealing,zhou2008metal,schulte2013bandgap} the various stacking configurations,\cite{ohta2007interlayer,siegel2013charge,bostwick2007quasiparticle,ohta2006controlling,coletti2013revealing} and the distinct adatom adsorptions.\cite{zhou2008metal,schulte2013bandgap} The AB-stacked graphite exhibits the 3D $\pi$ energy bands, with the bilayer and monolayer-like energy dispersions, respectilvey, at $k_x = 0$ and zone boundary of $k_x = 1$ (K and H points in the 3D first Brillouin zone).\cite{sugawara2006fermi,gruneis2008electron} The verified electronic structures of 2D few-layer graphenes include the Dirac-cone structure in single-layer system,\cite{ohta2007interlayer,siegel2013charge,bostwick2007quasiparticle} two pairs of parabolic bands in bilayer AB stacking,\cite{ohta2007interlayer,ohta2006controlling} the linear and parabolic bands in tri-layer ABA stacking,\cite{ohta2007interlayer,coletti2013revealing} the partially flat, sombrero-shaped and linear bands in tri-layer ABC stacking,\cite{coletti2013revealing} the red-shift Dirac cone in alkali-adsorbed graphenes,\cite{zhou2008metal} and the gap opening in graphene oxides.\cite{schulte2013bandgap} The 1D parabolic energy bands in graphene nanoribbons have been directly observed.\cite{ruffieux2012electronic} Up to now, the ARPES measurements on the unusual energy bands in alkali-adsorbed GNRs are absent. The further examinations are required for the occupied valence and conduction bands near the Fermi level, including the alkali-dominated conduction bands, the carbon-related conduction and valence bands, and the distribution-, concentration- and edge-dependent ones. They are useful in understanding the single $s-2p_z$ orbital hybridization in the alkali-C bond. Specifically, the experimental verifications on the existence of Fermi momenta or energy gap and the spin degeneracy of edge-localized energy bands can identify the spin configurations in zigzag GNRs.

The orbital hybridizations in chemical bondings, which are responsible for the rich electronic properties, are clearly evidenced by the spatial charge distributions. They could be characterized by the carrier density ($\rho$) and the difference of carrier density (${\Delta\rho}$). The latter is created by subtracting the pristine system from that of the adatom-adsorbed one. For a planar graphene nanoribbons, the parallel $2p_z$ orbitals and the planar ($2s$; $2p_x$; $2p_y$) orbitals can form the $\pi$ bondings and the $\sigma$ bondings, respectively, as shown by the dashed and the solid rectangles in Fig. 5(a). The $\sigma$ orbital hybridizations, with very high charge density between two carbon atoms, belong to covalent bonds. They keep the same under the alkali adatom adsorptions, corresponding to the zero $\Delta\rho$ in Figs. 5(f)-5(i). Also, the $\pi$ bondings could survive in alkali-adsorbed graphenes for any concentrations, distributions and edge structures, as illustrated in Figs. 5(b)-5(e). However, the modifications on them are observable through the charge variations between alkali and carbon atoms (the dashed rectangles in Figs. 5(f)-5(i)). Also shown is that $\Delta \rho$ is absent in neighboring carbon atoms, or the planar $\sigma$ bonding remains the same after alkali adsorption. These indicate that there only exist a single $s-2p_z$ orbital hybridization in the A-C bond.

DOS can exhibit many special structures due to the band-edge states in 1D energy dispersions, as clearly indicated in Figs. 6(a)-6(h). The asymmetric and symmetric peaks are, respectively, presented in the square-root and delta-function-like divergent forms, being associated with the parabolic and partially flat energy dispersions. Their intensities are proportional to the inverse of curvature and the dispersionless $k_x$-range. For a pristine armchair GNR, an energy gap, with zero DOS, is revealed between one pair of opposite-side anti-symmetric peaks (Fig. 6(a)). It almost keeps the same after alkali adsorption, but changes into an energy spacing between valence and conduction bands. The prominent peaks within $|E|\le\,3$ eV are dominated by $2p_z$ orbitals (red curve) and those of $E<-3$ eV by $2p_x+2p_y$ orbitals (green curve). There are minor modifications on the carbon-dependent peak structures under the alkali adsorption (Figs. 6(b) and 6(a)), i.e., a red shift of DOS could be observed from the change of $E_F$. Specifically, the adatoms can create pronounced asymmetric peaks near $E_F$, corresponding to the alkali-dominated occupied or unoccupied conduction bands. The energy and number of special structures depend on the distribution and concentration (Figs. 6(c) and 6(d)). Such peaks might merge with the carbon-dependent ones. The main differences between zigzag and armchair GNRs lie in the low-lying peak structures. The former can present the delta-function-like symmetric structures arising from the edge-localized energy bands, being sensitive to the spin configurations. The anti-ferromagnetic, ferromagnetic and non-magnetic zigzag GNRs, respectively, possess a pair of symmetric peaks (blue triangles in Figs. 6(e) and 6(g)), three peaks (Fig. 6(f)), and a merged peak (Fig. 6(h)). The peak intensities are obviously reduced for the spin-split energy bands (red circles in Fig. 6(f)).

The STS measurements, in which the differential tunneling conductance (dI/dV) is approximately proportional to DOS and directly reflects its special structures, could serve as an efficient way to confirm the contributions of carbon and alkali atoms. They have been successfully utilized to verify the diverse electronic properties in graphite,\cite{klusek1999investigations} few-layer graphenes,\cite{li2009scanning,luican2011single,li2010observation,cherkez2015van} carbon nanotubes,\cite{wilder1998electronic,odom1998atomic} and GNRs.\cite{tapaszto2008tailoring,huang2012spatially,sode2015electronic,chen2013tuning} The experimental identifications include the splitting $\pi$ and $\pi^\ast$ peaks at middle energy of $|E|\sim\,2$ eV and a finite DOS at $E_F$ in semi-metallic AB-stacked graphite, a V-shaped spectrum vanishing at the Dirac point in monolayer graphene,\cite{li2009scanning} the asymmetry-created peak structures in bilayer graphene,\cite{luican2011single,li2010observation,cherkez2015van} and the geometry-dependent energy gaps and the asymmetric peaks of 1D parabolic bands in carbon nanotubes and GNRs. The main features of electronic properties, including the energy spacing between valence and conduction bands, the alkali-dominated special structures close to $E_F$, a lot of carbon-induced asymmetric peaks at a whole energy range, and the prominent symmetric peaks of partially flat edge-localized bands with or without spin splitting could be further investigated with STS. The STS measurements on the low- and middle-energy peak structures are useful in identifying the single-orbital bonding and the specific spin configurations in alkali-adsorbed GNRs.

\section{Concluding Remarks}

The geometric structures, electronic and magnetic properties of alkali-adsorbed GNRs
are studied using the first-principles calculations. They are shown to be, respectively, determined by the adatom position, a single $s-2p_z$ orbital hybridization in A-C bond, and the spin configurations at two edges. The critical chemical bonding is responsible for the feature-rich electronic properties, including the occupied carbon- and alkali-dominated conduction bands near $E_F$, and a linear relation between the linear free electron density and the adatom concentration. Any alkali atoms contribute the outmost s orbitals to become free carriers in adsorbed systems. The creation of high free carrier density indicates that alkali-adsorbed GNRs might have potential applications in nanoelectronic devices.\cite{obradovic2006analysis,wang2008room} Moreover, the single-edge adsorption causes ZGNRs to exhibit the spin-split metallic energy bands. They could be considered as promising materials for future applications in spintronic devices.\cite{saffarzadeh2011spin}

The essential properties are diversified by the cooperative or competitive relations among quantum confinement, A-C interaction, and spin configuration. GNRs keep a planar structure after alkali adsorption, indicating the unchanged $\sigma$ bondings of carbon atoms. The optimal position of adatom is located at the hollow site, in which its height grows with the atomic number. The predicted geometric structure could be verified by the STM measurements.\cite{wilder1998electronic,odom1998atomic} GNRs change from semiconductors to metals during the gradual increase of adsorption concentration. The alkali-adsorbed GNRs possess the well-extended $\pi$ bonding, while there exist charge variations between alkali and carbon atoms. The former is the main reason for the minor modifications on the carbon-dominated valence bands. The latter arises from the $s-2p_z$ hybridization so that the alkali- and carbon-dependent conduction bands rely on the kind, distribution, and concentration of adatom. Also, the edge structure plays a critical role in magnetic and electronic properties. Zigzag systems can present the anti-ferromagnetic ordering between two edges, the ferromagnetic configuration only at one edge, or the non-magnetism, depending on whether adatoms are close to the boundaries. They have the partially flat edge-localized energy bands near $E_F$; furthermore, both spin-up and spin-down states are split under the single-edge adsorption. The distinct free carrier density in the splitting energy bands could create the spin-polarized current in transport measurements. The ARPES measurements\cite{sugawara2006fermi,gruneis2008electron,ohta2007interlayer,siegel2013charge,bostwick2007quasiparticle,ohta2006controlling,coletti2013revealing,schulte2013bandgap,zhou2008metal} are suitable for the identification of diverse energy bands. The 1D energy dispersions are directly reflected in DOS as a lot of special structures. The anti-symmetric and symmetric peaks, respectively, correspond to the parabolic and partially flat energy bands. The STS measurements on the alkali-dominated anti-symmetric peaks near $E_F$ could be utilized to examine the critical chemical bonding. The distinct spin configurations in alkali-adsorbed zigzag GNRs are distinguishable by measuring the number, energy, and intensity of low-lying symmetric peaks.

This work shows that the first-principles calculations, combined with the critical orbital hybridizations in adatom-carbon and carbon-carbon bonds and the specific spin arrangements, are useful in exploring the orbital- and spin-dominated essential properties. For example, band structure, free carrier density, and DOS are determined by which kinds of atomic orbitals and spin configurations These could be further generalized to the layered condensed-matter systems, with the nano-scaled thickness and the unique lattice symmetries. In addition to graphene, the emergent layered materials include silicene, germanene, tinene, phosphorene, MoS$_2$ and so on. Whether the alkali adsorptions on these semiconducting nanoribbon systems can create the high free carrier density deserves a thorough and systematic study. The complicated relations among quantum confinement, geometric symmetry, multi-orbital hybridizations in chemical bonds, and spin-orbital interactions, being responsible for the essential properties, need to be investigated in detail.

\begin{acknowledgement}
This work was supported by the Nation Science Council of Taiwan (Grant No. NSC 105-2112-M-006-002-MY3). We also thank the National Center for High-performance Computing (NCHC) for computer facilities.

\end{acknowledgement}

\newpage
\begin{table}
  \caption{ Bond lengths, heights, and free electron densities of $N_A=12$ alkali-adsorbed GNRs}
  \label{tbl:notes}
  \begin{tabular}{ccccccc}
     \hline
    \hline
    N$_A$= 12 & nearest C-C  & bond length & height  & $\lambda$ ($10^7$ $ e/cm$)& $\lambda$ /adatom & Bader charge \\
    &(\AA{}) & A-C (\AA{}) &  (\AA{}) & & concentration (e) & transfer (e) \\

    \hline
    Li & 1.384 & 2.23 & 1.75 & 2.31 & 0.995 & 0.75 \\
    Na & 1.377 & 2.27 & 2.46 & 2.35 & 1.011 & 0.44 \\
    K & 1.375 & 3.21 & 2.81 & 2.31 & 0.995 & 0.45 \\
    Rb& 1.374 & 3.36 & 2.96 & 2.36 & 1.013 & 0.47 \\
    Cs & 1.374 & 3.37 & 3.06 & 2.31 & 0.995 & 0.54 \\

    \hline
    \hline

\end{tabular}

\end{table}

\begin{table}
  \caption{ Free electron densities for various numbers and positions of adatoms in $N_A=12$ and $N_z=8$ alkali-adsorbed GNRs}
  \label{tbl:notes}
  \begin{tabular}{ccccccc}
     \hline
    \hline
    GNR systems & configurations &  $\lambda$ ($10^7$ $ e/cm$)& $\lambda$ /adatom & Bader charge \\
    & &  & concentration (e)  & transfer (e) \\

    \hline
    				 	&  (3,7)$_{single}$ & 4.67  & 1.008 & 0.69  \\
     					 & (3,{\color{red}{7}})$_{double}$ & 4.62  & 0.989 & 0.75   \\
    Armchair	 & (1,10)$_{single}$  & 4.68  & 1.008 & 0.74   \\
    N$_A$= 12 & (1,{\color{red}{10}})$_{double}$  & 4.67 & 1.008 & 0.74   \\
    					& (1,{\color{red}{4}},7,{\color{red}{10}})$_{double}$  & 9.24 & 0.994 & 0.71   \\
     					& (1,{\color{red}{2}},5,{\color{red}{6}},9,{\color{red}{10}})$_{double}$ & 13.96 & 1.007 & 0.70  \\

    \hline
    			  	 	&  (1)$_{single}$ & 2.02  & 1.001 & 0.88  \\
     Zigzag		 	& (7)$_{single}$ & 2.03  & 1.009 & 0.89   \\
   	N$_Z$= 8 	& (1,13)$_{single}$  & 4.10  & 1.015 & 0.88  \\
   						& (1,{\color{red}{13}})$_{double}$  & 4.02 & 1.001 & 0.88   \\

    \hline
	\hline
\end{tabular}

\end{table}

\begin{figure}[h]
\centering
  \includegraphics[width=0.8\linewidth]{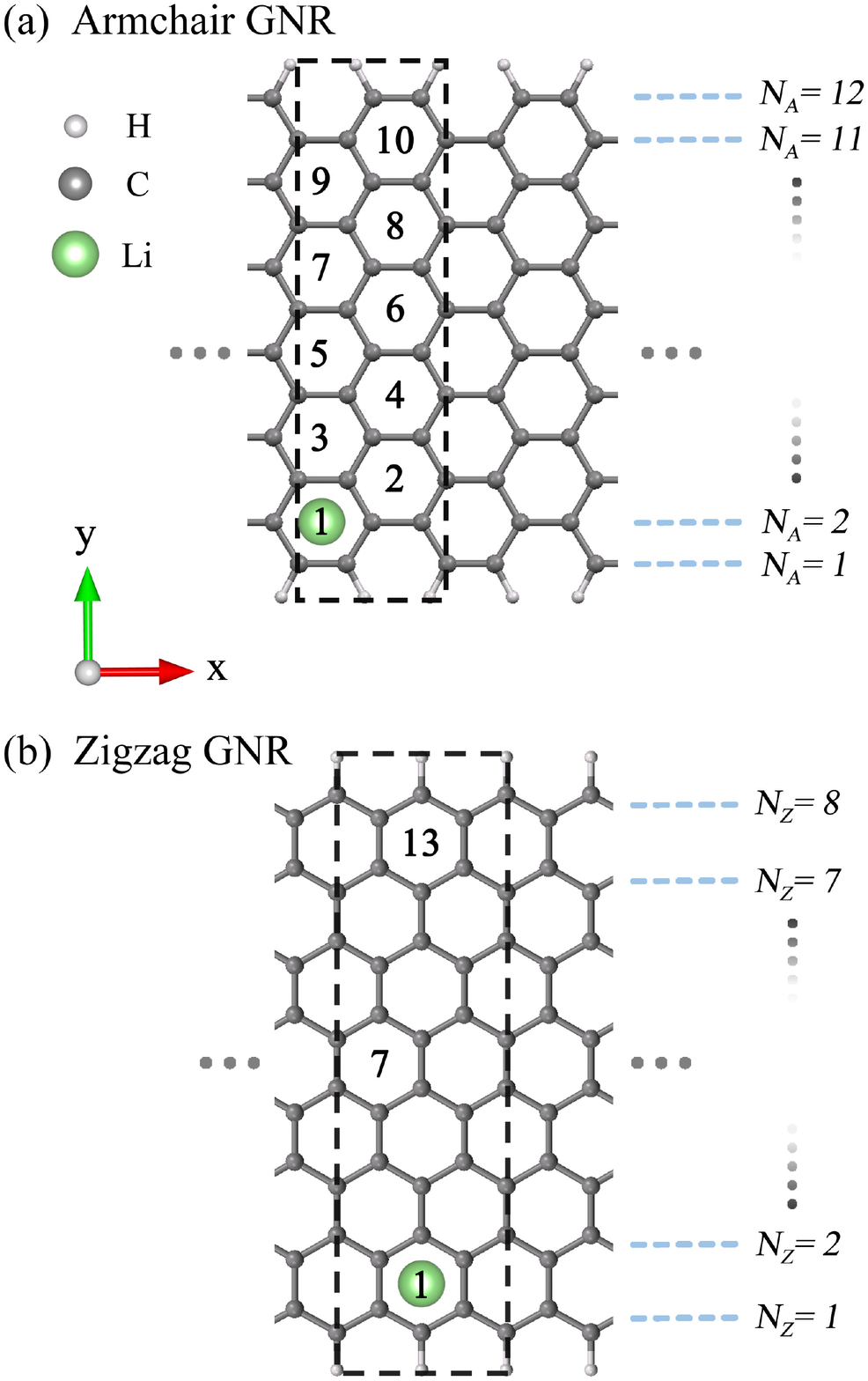}
  \caption{geometric structures of alkali-adsorbed GNRs for (a) $N_A=12$ armchair and (b) $N_Z=8$ zigzag systems. The dashed rectangles represent unit cells used in the calculations. The lattice constants are, respectively, $a=3b$ and $a=2\sqrt 3\,b$ for armchair and zigzag GNRs. Numbers inside hexagons denote the adatom adsorption positions.}
  \label{fgr:1}
\end{figure}

\begin{figure}[h]
\centering
  \includegraphics[width=0.8\linewidth]{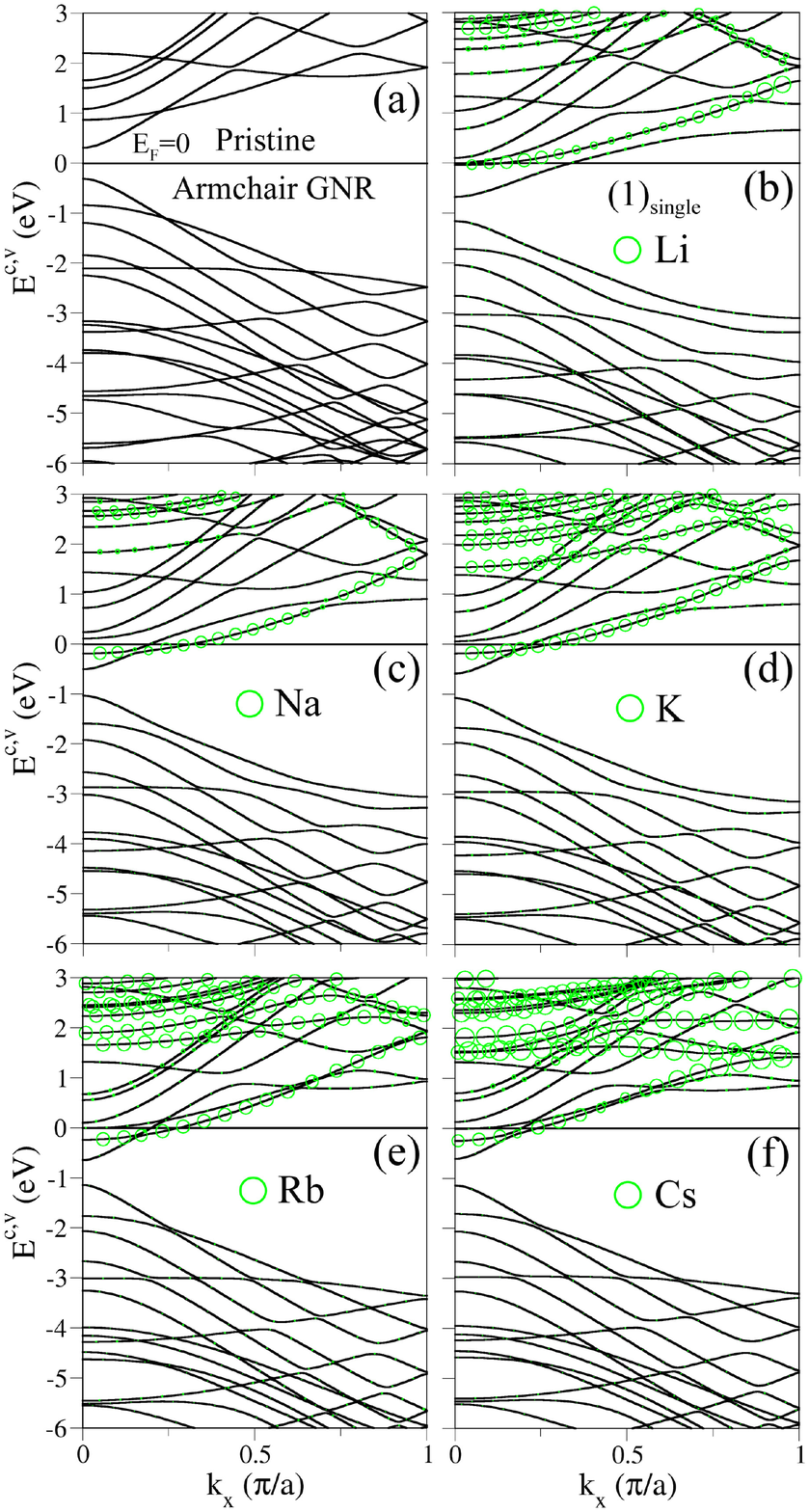}
  \caption{Band structures of $N_A=12$ armchair systems for (a) a pristine GNR, and the (b) Li-, (c) Na-, (d) K-, (e) Rb- and (f) Cs-adsorbed GNRs with an adtom at edge. Green circles represent the contributions of alkali adatoms.}
  \label{fgr:2}
\end{figure}

\begin{figure}[h]
\centering
  \includegraphics[width=0.8\linewidth]{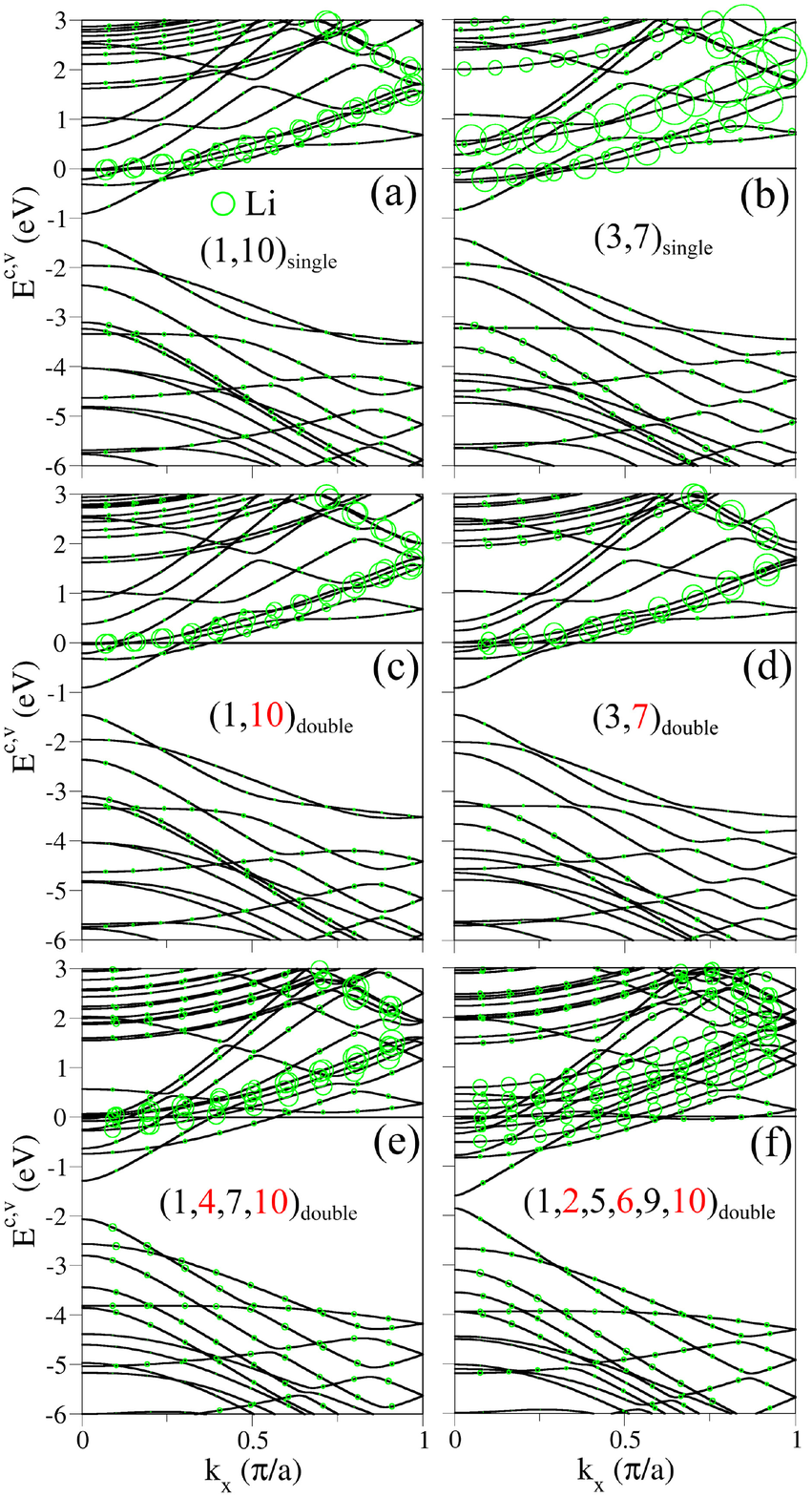}
  \caption{Similar plot as Fig. 2, but shown for the Li-adsorbed GNRs. (a) and (b), respectively, correspond to the single-side adsorptions of adatoms at (1,10)$_{single}$ and (3,7)$_{single}$. The double-side adsorptions are shown for two adtoms at (c) (1,\textcolor{red}{10})$_{double}$ \& (d) (3,\textcolor{red}{7})$_{double}$, four adatoms at (1,\textcolor{red}{4},7,\textcolor{red}{10})$_{double}$, and six adatoms at (1,\textcolor{red}{2},5,\textcolor{red}{6},9,\textcolor{red}{10})$_{double}$. Numbers correspond to the adatom adsorption position. Two colors (black \& red) denote two distinct sides.}
  \label{fgr:3}
\end{figure}

\begin{figure}[h]
\centering
  \includegraphics[width=0.8\linewidth]{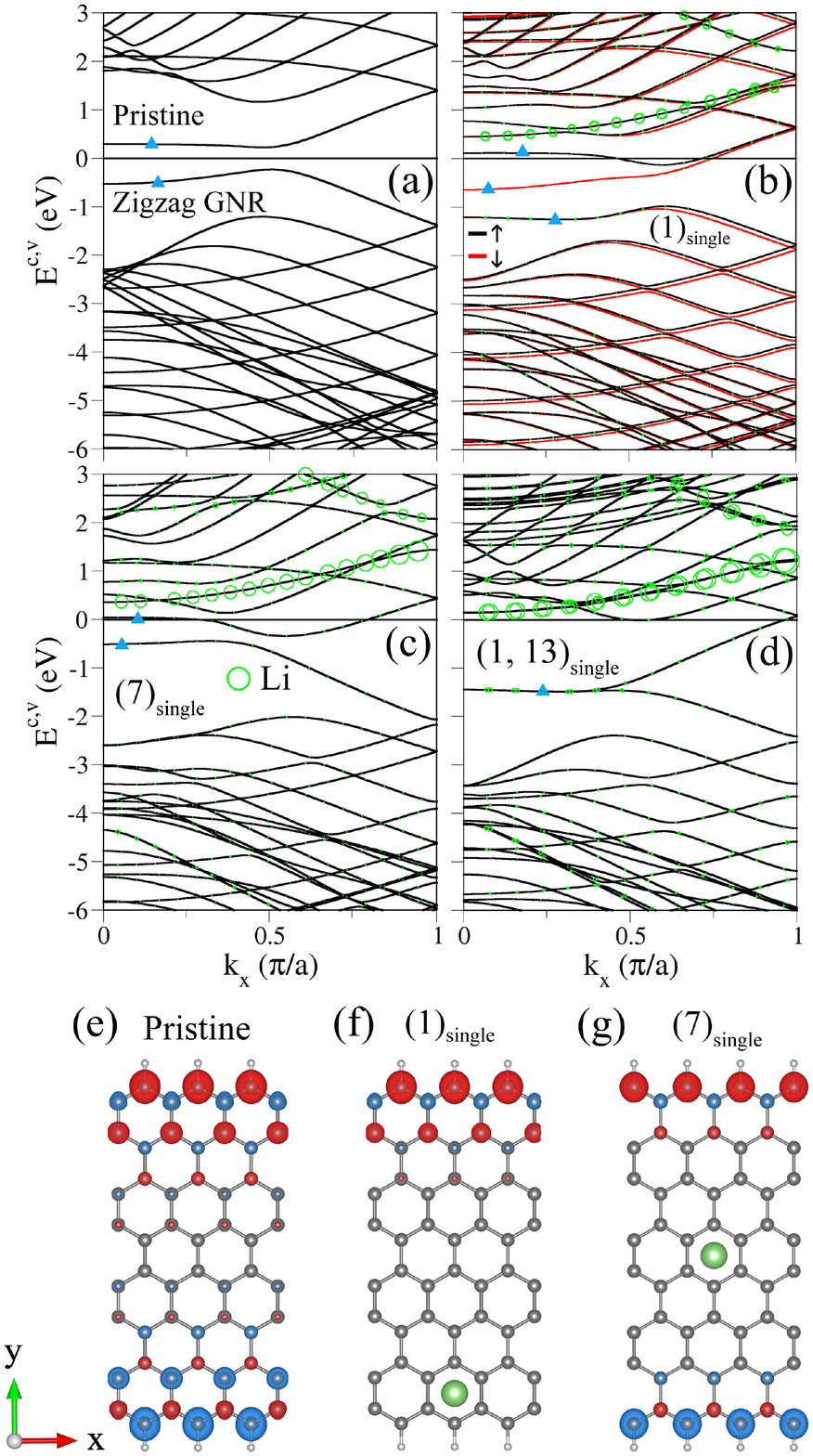}
  \caption{Band structures of $N_Z=8$ zigzag systems for (a) a pristine GNR and the Li adsorptions situated at (b) a single edge, (c) a ribbon center and (d) two edges of single side. The spin configurations of (a), (b) and (c) are indicated in (e), (f) and (g), respectively. Blue and red circles, respectively, represent spin-up and spin-down arrangements.}
  \label{fgr:4}
\end{figure}

\begin{figure}[h]
\centering
  \includegraphics[width=0.8\linewidth]{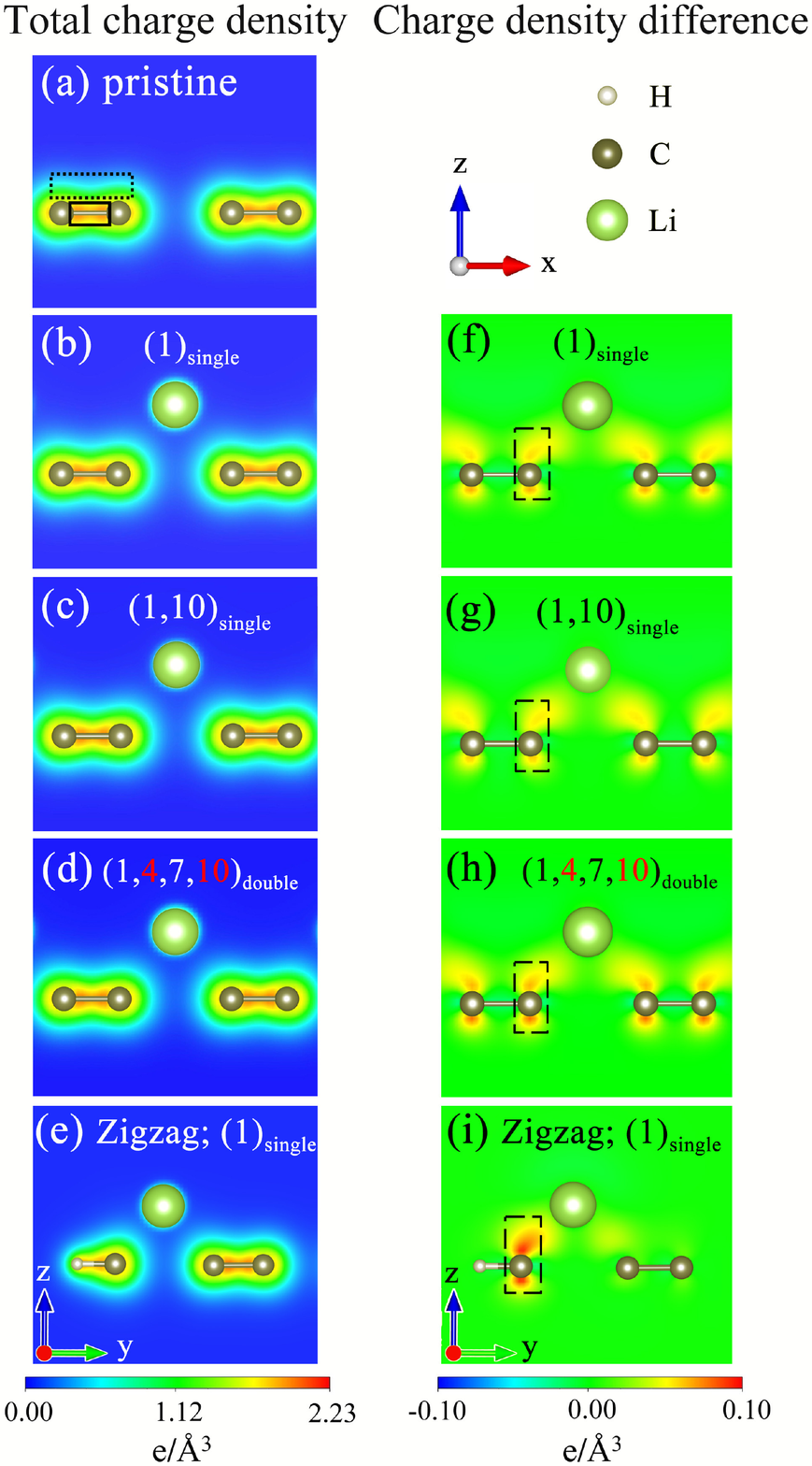}
  \caption{The spatial charge distributions of $N_A=12$ armchair systems for (a) a pristine GNR, (b) an adatom at edge (1)$_{single}$, (c) two adatoms at (1,10)$_{single}$ and (d) four adatoms at (1,\textcolor{red}{4},7,\textcolor{red}{10})$_{double}$. The $\pi$ and $\sigma$ bondings are, respectively, enclosed by the dashed and solid rectangles. The charge density differences, corresponding to (b), (c) and (d), are revealed in (f), (g) and (h), respectively. Also shown in (e) and (i) are those of $N_z=8$ zigzag system with an adatom.}
  \label{fgr:5}
\end{figure}

\begin{figure}[h]
\centering
  \includegraphics[width=0.8\linewidth]{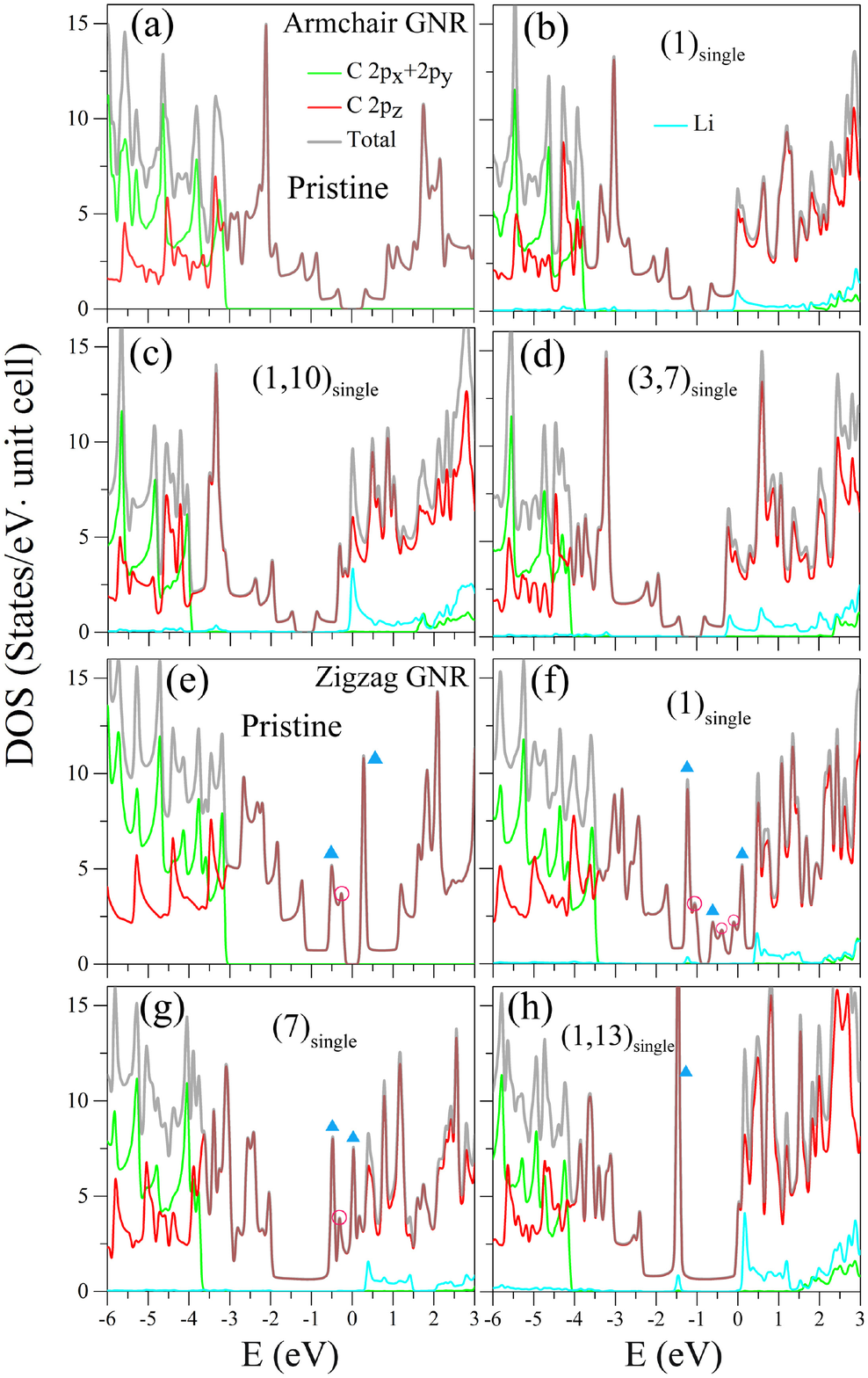}
  \caption{The total and orbital-projected DOSs of $N_A=12$ armchair systems for (a) a pristine GNR, (b) an adatom at edge (1)$_{single}$, and the double-side adsorptions of (c) two adatoms at (1,10)$_{single}$ and (d) (3,7)$_{single}$. Those of $N_Z=8$ zigzag systems are shown for (e) a pristine GNR, and the Li adsorptions situated at (f) (1)$_{single}$, (g) a ribbon center (7)$_{single}$ and (h) two edges (1,13)$_{single}$.}
  \label{fgr:6}
\end{figure}


\newpage
\bibliography{Ribbon-Xavier}

\providecommand{\latin}[1]{#1}
\providecommand*\mcitethebibliography{\thebibliography}
\csname @ifundefined\endcsname{endmcitethebibliography}
  {\let\endmcitethebibliography\endthebibliography}{}
\begin{mcitethebibliography}{89}
\providecommand*\natexlab[1]{#1}
\providecommand*\mciteSetBstSublistMode[1]{}
\providecommand*\mciteSetBstMaxWidthForm[2]{}
\providecommand*\mciteBstWouldAddEndPuncttrue
  {\def\EndOfBibitem{\unskip.}}
\providecommand*\mciteBstWouldAddEndPunctfalse
  {\let\EndOfBibitem\relax}
\providecommand*\mciteSetBstMidEndSepPunct[3]{}
\providecommand*\mciteSetBstSublistLabelBeginEnd[3]{}
\providecommand*\EndOfBibitem{}
\mciteSetBstSublistMode{f}
\mciteSetBstMaxWidthForm{subitem}{(\alph{mcitesubitemcount})}
\mciteSetBstSublistLabelBeginEnd
  {\mcitemaxwidthsubitemform\space}
  {\relax}
  {\relax}

\bibitem[Tuinstra and Koenig(1970)Tuinstra, and Koenig]{tuinstra1970raman}
Tuinstra,~F.; Koenig,~J.~L. \emph{The Journal of Chemical Physics}
  \textbf{1970}, \emph{53}, 1126--1130\relax
\mciteBstWouldAddEndPuncttrue
\mciteSetBstMidEndSepPunct{\mcitedefaultmidpunct}
{\mcitedefaultendpunct}{\mcitedefaultseppunct}\relax
\EndOfBibitem
\bibitem[Huang \latin{et~al.}(1997)Huang, Lin, and Chuu]{huang1997collective}
Huang,~C.; Lin,~M.; Chuu,~D. \emph{Solid state communications} \textbf{1997},
  \emph{103}, 603--606\relax
\mciteBstWouldAddEndPuncttrue
\mciteSetBstMidEndSepPunct{\mcitedefaultmidpunct}
{\mcitedefaultendpunct}{\mcitedefaultseppunct}\relax
\EndOfBibitem
\bibitem[Novoselov \latin{et~al.}(2004)Novoselov, Geim, Morozov, Jiang, Zhang,
  Dubonos, , Grigorieva, and Firsov]{novoselov2004electric}
Novoselov,~K.~S.; Geim,~A.~K.; Morozov,~S.; Jiang,~D.; Zhang,~Y.; Dubonos,~S.;
  ; Grigorieva,~I.; Firsov,~A. \emph{science} \textbf{2004}, \emph{306},
  666--669\relax
\mciteBstWouldAddEndPuncttrue
\mciteSetBstMidEndSepPunct{\mcitedefaultmidpunct}
{\mcitedefaultendpunct}{\mcitedefaultseppunct}\relax
\EndOfBibitem
\bibitem[Novoselov \latin{et~al.}(2005)Novoselov, Geim, Morozov, Jiang,
  Katsnelson, Grigorieva, Dubonos, and Firsov]{novoselov2005two}
Novoselov,~K.; Geim,~A.~K.; Morozov,~S.; Jiang,~D.; Katsnelson,~M.;
  Grigorieva,~I.; Dubonos,~S.; Firsov,~A. \emph{nature} \textbf{2005},
  \emph{438}, 197--200\relax
\mciteBstWouldAddEndPuncttrue
\mciteSetBstMidEndSepPunct{\mcitedefaultmidpunct}
{\mcitedefaultendpunct}{\mcitedefaultseppunct}\relax
\EndOfBibitem
\bibitem[Son \latin{et~al.}(2006)Son, Cohen, and Louie]{son2006half}
Son,~Y.~W.; Cohen,~M.~L.; Louie,~S.~G. \emph{Nature} \textbf{2006}, \emph{444},
  347--349\relax
\mciteBstWouldAddEndPuncttrue
\mciteSetBstMidEndSepPunct{\mcitedefaultmidpunct}
{\mcitedefaultendpunct}{\mcitedefaultseppunct}\relax
\EndOfBibitem
\bibitem[Li \latin{et~al.}(2008)Li, Wang, Zhang, Lee, and
  Dai]{li2008chemically}
Li,~X.; Wang,~X.; Zhang,~L.; Lee,~S.; Dai,~H. \emph{Science} \textbf{2008},
  \emph{319}, 1229--1232\relax
\mciteBstWouldAddEndPuncttrue
\mciteSetBstMidEndSepPunct{\mcitedefaultmidpunct}
{\mcitedefaultendpunct}{\mcitedefaultseppunct}\relax
\EndOfBibitem
\bibitem[Iijima(1991)]{iijima1991helical}
Iijima,~S. \emph{nature} \textbf{1991}, \emph{354}, 56--58\relax
\mciteBstWouldAddEndPuncttrue
\mciteSetBstMidEndSepPunct{\mcitedefaultmidpunct}
{\mcitedefaultendpunct}{\mcitedefaultseppunct}\relax
\EndOfBibitem
\bibitem[Iijima and Ichihashi(1993)Iijima, and Ichihashi]{iijima1993single}
Iijima,~S.; Ichihashi,~T. \emph{Nature} \textbf{1993}, \emph{363},
  603--605\relax
\mciteBstWouldAddEndPuncttrue
\mciteSetBstMidEndSepPunct{\mcitedefaultmidpunct}
{\mcitedefaultendpunct}{\mcitedefaultseppunct}\relax
\EndOfBibitem
\bibitem[Datta \latin{et~al.}(2008)Datta, Strachan, Khamis, and
  Johnson]{datta2008crystallographic}
Datta,~S.~S.; Strachan,~D.~R.; Khamis,~S.~M.; Johnson,~A.~C. \emph{Nano
  letters} \textbf{2008}, \emph{8}, 1912--1915\relax
\mciteBstWouldAddEndPuncttrue
\mciteSetBstMidEndSepPunct{\mcitedefaultmidpunct}
{\mcitedefaultendpunct}{\mcitedefaultseppunct}\relax
\EndOfBibitem
\bibitem[Ci \latin{et~al.}(2008)Ci, Xu, Wang, Gao, Ding, Kelly, Yakobson, and
  Ajayan]{ci2008controlled}
Ci,~L.; Xu,~Z.; Wang,~L.; Gao,~W.; Ding,~F.; Kelly,~K.~F.; Yakobson,~B.~I.;
  Ajayan,~P.~M. \emph{Nano Research} \textbf{2008}, \emph{1}, 116--122\relax
\mciteBstWouldAddEndPuncttrue
\mciteSetBstMidEndSepPunct{\mcitedefaultmidpunct}
{\mcitedefaultendpunct}{\mcitedefaultseppunct}\relax
\EndOfBibitem
\bibitem[McAllister \latin{et~al.}(2007)McAllister, Li, Adamson, Schniepp,
  Abdala, Liu, Herrera-Alonso, Milius, Car, and
  Prud'homme]{mcallister2007single}
McAllister,~M.~J.; Li,~J.~L.; Adamson,~D.~H.; Schniepp,~H.~C.; Abdala,~A.~A.;
  Liu,~J.; Herrera-Alonso,~M.; Milius,~D.~L.; Car,~R.; Prud'homme,~R.~K.
  \emph{Chemistry of materials} \textbf{2007}, \emph{19}, 4396--4404\relax
\mciteBstWouldAddEndPuncttrue
\mciteSetBstMidEndSepPunct{\mcitedefaultmidpunct}
{\mcitedefaultendpunct}{\mcitedefaultseppunct}\relax
\EndOfBibitem
\bibitem[Fujii and Enoki(2010)Fujii, and Enoki]{fujii2010cutting}
Fujii,~S.; Enoki,~T. \emph{Journal of the American Chemical Society}
  \textbf{2010}, \emph{132}, 10034--10041\relax
\mciteBstWouldAddEndPuncttrue
\mciteSetBstMidEndSepPunct{\mcitedefaultmidpunct}
{\mcitedefaultendpunct}{\mcitedefaultseppunct}\relax
\EndOfBibitem
\bibitem[Han \latin{et~al.}(2007)Han, {\"O}zyilmaz, Zhang, and
  Kim]{han2007energy}
Han,~M.~Y.; {\"O}zyilmaz,~B.; Zhang,~Y.; Kim,~P. \emph{Physical review letters}
  \textbf{2007}, \emph{98}, 206805\relax
\mciteBstWouldAddEndPuncttrue
\mciteSetBstMidEndSepPunct{\mcitedefaultmidpunct}
{\mcitedefaultendpunct}{\mcitedefaultseppunct}\relax
\EndOfBibitem
\bibitem[Chen \latin{et~al.}(2007)Chen, Lin, Rooks, and
  Avouris]{chen2007graphene}
Chen,~Z.; Lin,~Y.~M.; Rooks,~M.~J.; Avouris,~P. \emph{Physica E:
  Low-dimensional Systems and Nanostructures} \textbf{2007}, \emph{40},
  228--232\relax
\mciteBstWouldAddEndPuncttrue
\mciteSetBstMidEndSepPunct{\mcitedefaultmidpunct}
{\mcitedefaultendpunct}{\mcitedefaultseppunct}\relax
\EndOfBibitem
\bibitem[Wang \latin{et~al.}(2008)Wang, Ouyang, Li, Wang, Guo, and
  Dai]{wang2008room}
Wang,~X.; Ouyang,~Y.; Li,~X.; Wang,~H.; Guo,~J.; Dai,~H. \emph{Physical review
  letters} \textbf{2008}, \emph{100}, 206803\relax
\mciteBstWouldAddEndPuncttrue
\mciteSetBstMidEndSepPunct{\mcitedefaultmidpunct}
{\mcitedefaultendpunct}{\mcitedefaultseppunct}\relax
\EndOfBibitem
\bibitem[Ruffieux \latin{et~al.}(2012)Ruffieux, Cai, Plumb, Patthey, Prezzi,
  Ferretti, Molinari, Feng, Mu?llen, and Pignedoli]{ruffieux2012electronic}
Ruffieux,~P.; Cai,~J.; Plumb,~N.~C.; Patthey,~L.; Prezzi,~D.; Ferretti,~A.;
  Molinari,~E.; Feng,~X.; Mu?llen,~K.; Pignedoli,~C.~A. \emph{Acs Nano}
  \textbf{2012}, \emph{6}, 6930--6935\relax
\mciteBstWouldAddEndPuncttrue
\mciteSetBstMidEndSepPunct{\mcitedefaultmidpunct}
{\mcitedefaultendpunct}{\mcitedefaultseppunct}\relax
\EndOfBibitem
\bibitem[Huang \latin{et~al.}(2012)Huang, Wei, Sun, Wong, Feng, Neto, and
  Wee]{huang2012spatially}
Huang,~H.; Wei,~D.; Sun,~J.; Wong,~S.~L.; Feng,~Y.~P.; Neto,~A.~C.; Wee,~A.
  T.~S. \emph{Scientific reports} \textbf{2012}, \emph{2}, 983\relax
\mciteBstWouldAddEndPuncttrue
\mciteSetBstMidEndSepPunct{\mcitedefaultmidpunct}
{\mcitedefaultendpunct}{\mcitedefaultseppunct}\relax
\EndOfBibitem
\bibitem[Kosynkin \latin{et~al.}(2009)Kosynkin, Higginbotham, Sinitskii,
  Lomeda, Dimiev, Price, and Tour]{kosynkin2009longitudinal}
Kosynkin,~D.~V.; Higginbotham,~A.~L.; Sinitskii,~A.; Lomeda,~J.~R.; Dimiev,~A.;
  Price,~B.~K.; Tour,~J.~M. \emph{Nature} \textbf{2009}, \emph{458},
  872--876\relax
\mciteBstWouldAddEndPuncttrue
\mciteSetBstMidEndSepPunct{\mcitedefaultmidpunct}
{\mcitedefaultendpunct}{\mcitedefaultseppunct}\relax
\EndOfBibitem
\bibitem[Cataldo \latin{et~al.}(2010)Cataldo, Compagnini, Patan{\'e}, Ursini,
  Angelini, Ribic, Margaritondo, Cricenti, Palleschi, and
  Valentini]{cataldo2010graphene}
Cataldo,~F.; Compagnini,~G.; Patan{\'e},~G.; Ursini,~O.; Angelini,~G.;
  Ribic,~P.~R.; Margaritondo,~G.; Cricenti,~A.; Palleschi,~G.; Valentini,~F.
  \emph{Carbon} \textbf{2010}, \emph{48}, 2596--2602\relax
\mciteBstWouldAddEndPuncttrue
\mciteSetBstMidEndSepPunct{\mcitedefaultmidpunct}
{\mcitedefaultendpunct}{\mcitedefaultseppunct}\relax
\EndOfBibitem
\bibitem[Kumar \latin{et~al.}(2011)Kumar, Panchakarla, and Rao]{kumar2011laser}
Kumar,~P.; Panchakarla,~L.; Rao,~C. \emph{Nanoscale} \textbf{2011}, \emph{3},
  2127--2129\relax
\mciteBstWouldAddEndPuncttrue
\mciteSetBstMidEndSepPunct{\mcitedefaultmidpunct}
{\mcitedefaultendpunct}{\mcitedefaultseppunct}\relax
\EndOfBibitem
\bibitem[El\'{\i}as \latin{et~al.}(2009)El\'{\i}as, Botello-M{\'e}ndez,
  Meneses-Rodr\'{\i}guez, Jehov\'{a}~Gonz\'{a}lez, Ram\'{\i}rez-Gonz\'{a}lez,
  Ci, Munoz-Sandoval, Ajayan, Terrones, and Terrones]{elias2009longitudinal}
El\'{\i}as,~A.~L.; Botello-M{\'e}ndez,~A.~R.; Meneses-Rodr\'{\i}guez,~D.;
  Jehov\'{a}~Gonz\'{a}lez,~V.; Ram\'{\i}rez-Gonz\'{a}lez,~D.; Ci,~L.;
  Munoz-Sandoval,~E.; Ajayan,~P.~M.; Terrones,~H.; Terrones,~M. \emph{Nano
  letters} \textbf{2009}, \emph{10}, 366--372\relax
\mciteBstWouldAddEndPuncttrue
\mciteSetBstMidEndSepPunct{\mcitedefaultmidpunct}
{\mcitedefaultendpunct}{\mcitedefaultseppunct}\relax
\EndOfBibitem
\bibitem[Parashar \latin{et~al.}(2011)Parashar, Bhandari, Srivastava, Jariwala,
  and Srivastava]{parashar2011single}
Parashar,~U.~K.; Bhandari,~S.; Srivastava,~R.~K.; Jariwala,~D.; Srivastava,~A.
  \emph{Nanoscale} \textbf{2011}, \emph{3}, 3876--3882\relax
\mciteBstWouldAddEndPuncttrue
\mciteSetBstMidEndSepPunct{\mcitedefaultmidpunct}
{\mcitedefaultendpunct}{\mcitedefaultseppunct}\relax
\EndOfBibitem
\bibitem[Jiao \latin{et~al.}(2009)Jiao, Zhang, Wang, Diankov, and
  Dai]{jiao2009narrow}
Jiao,~L.; Zhang,~L.; Wang,~X.; Diankov,~G.; Dai,~H. \emph{Nature}
  \textbf{2009}, \emph{458}, 877--880\relax
\mciteBstWouldAddEndPuncttrue
\mciteSetBstMidEndSepPunct{\mcitedefaultmidpunct}
{\mcitedefaultendpunct}{\mcitedefaultseppunct}\relax
\EndOfBibitem
\bibitem[Jiao \latin{et~al.}(2010)Jiao, Zhang, Ding, Liu, and
  Dai]{jiao2010aligned}
Jiao,~L.; Zhang,~L.; Ding,~L.; Liu,~J.; Dai,~H. \emph{Nano Research}
  \textbf{2010}, \emph{3}, 387--394\relax
\mciteBstWouldAddEndPuncttrue
\mciteSetBstMidEndSepPunct{\mcitedefaultmidpunct}
{\mcitedefaultendpunct}{\mcitedefaultseppunct}\relax
\EndOfBibitem
\bibitem[Talyzin \latin{et~al.}(2011)Talyzin, Luzan, Anoshkin, Nasibulin,
  Jiang, Kauppinen, Mikoushkin, Shnitov, Marchenko, and
  Noreus]{talyzin2011hydrogenation}
Talyzin,~A.~V.; Luzan,~S.; Anoshkin,~I.~V.; Nasibulin,~A.~G.; Jiang,~H.;
  Kauppinen,~E.~I.; Mikoushkin,~V.~M.; Shnitov,~V.~V.; Marchenko,~D.~E.;
  Noreus,~D. \emph{ACS nano} \textbf{2011}, \emph{5}, 5132--5140\relax
\mciteBstWouldAddEndPuncttrue
\mciteSetBstMidEndSepPunct{\mcitedefaultmidpunct}
{\mcitedefaultendpunct}{\mcitedefaultseppunct}\relax
\EndOfBibitem
\bibitem[Paiva \latin{et~al.}(2010)Paiva, Xu, Fernanda~Proen{\c{c}}a, Novais,
  L{\ae}gsgaard, and Besenbacher]{paiva2010unzipping}
Paiva,~M.~C.; Xu,~W.; Fernanda~Proen{\c{c}}a,~M.; Novais,~R.~M.;
  L{\ae}gsgaard,~E.; Besenbacher,~F. \emph{Nano letters} \textbf{2010},
  \emph{10}, 1764--1768\relax
\mciteBstWouldAddEndPuncttrue
\mciteSetBstMidEndSepPunct{\mcitedefaultmidpunct}
{\mcitedefaultendpunct}{\mcitedefaultseppunct}\relax
\EndOfBibitem
\bibitem[Kim \latin{et~al.}(2010)Kim, Sussman, and Zettl]{kim2010graphene}
Kim,~K.; Sussman,~A.; Zettl,~A. \emph{ACS nano} \textbf{2010}, \emph{4},
  3\relax
\mciteBstWouldAddEndPuncttrue
\mciteSetBstMidEndSepPunct{\mcitedefaultmidpunct}
{\mcitedefaultendpunct}{\mcitedefaultseppunct}\relax
\EndOfBibitem
\bibitem[Cano-M{\'a}rquez \latin{et~al.}(2009)Cano-M{\'a}rquez,
  Rodr{\'\i}guez-Mac{\'\i}as, Campos-Delgado, Espinosa-Gonz{\'a}lez,
  Trist{\'a}n-L{\'o}pez, Ram{\'\i}rez-Gonz{\'a}lez, Cullen, Smith, Terrones,
  and Vega-Cant{\'u}]{cano2009ex}
Cano-M{\'a}rquez,~A.~G.; Rodr{\'\i}guez-Mac{\'\i}as,~F.~J.; Campos-Delgado,~J.;
  Espinosa-Gonz{\'a}lez,~C.~G.; Trist{\'a}n-L{\'o}pez,~F.;
  Ram{\'\i}rez-Gonz{\'a}lez,~D.; Cullen,~D.~A.; Smith,~D.~J.; Terrones,~M.;
  Vega-Cant{\'u},~Y.~I. \emph{Nano letters} \textbf{2009}, \emph{9},
  1527--1533\relax
\mciteBstWouldAddEndPuncttrue
\mciteSetBstMidEndSepPunct{\mcitedefaultmidpunct}
{\mcitedefaultendpunct}{\mcitedefaultseppunct}\relax
\EndOfBibitem
\bibitem[Shinde \latin{et~al.}(2011)Shinde, Debgupta, Kushwaha, Aslam, and
  Pillai]{shinde2011electrochemical}
Shinde,~D.~B.; Debgupta,~J.; Kushwaha,~A.; Aslam,~M.; Pillai,~V.~K.
  \emph{Journal of the American Chemical Society} \textbf{2011}, \emph{133},
  4168--4171\relax
\mciteBstWouldAddEndPuncttrue
\mciteSetBstMidEndSepPunct{\mcitedefaultmidpunct}
{\mcitedefaultendpunct}{\mcitedefaultseppunct}\relax
\EndOfBibitem
\bibitem[Campos-Delgado \latin{et~al.}(2008)Campos-Delgado, Romo-Herrera, Jia,
  Cullen, Muramatsu, Kim, Hayashi, Ren, Smith, and Okuno]{campos2008bulk}
Campos-Delgado,~J.; Romo-Herrera,~J.~M.; Jia,~X.; Cullen,~D.~A.; Muramatsu,~H.;
  Kim,~Y.~A.; Hayashi,~T.; Ren,~Z.; Smith,~D.~J.; Okuno,~Y. \emph{Nano letters}
  \textbf{2008}, \emph{8}, 2773--2778\relax
\mciteBstWouldAddEndPuncttrue
\mciteSetBstMidEndSepPunct{\mcitedefaultmidpunct}
{\mcitedefaultendpunct}{\mcitedefaultseppunct}\relax
\EndOfBibitem
\bibitem[Yang \latin{et~al.}(2008)Yang, Dou, Rouhanipour, Zhi, R{\"a}der, and
  M{\"u}llen]{yang2008two}
Yang,~X.; Dou,~X.; Rouhanipour,~A.; Zhi,~L.; R{\"a}der,~H.~J.; M{\"u}llen,~K.
  \emph{Journal of the American Chemical Society} \textbf{2008}, \emph{130},
  4216--4217\relax
\mciteBstWouldAddEndPuncttrue
\mciteSetBstMidEndSepPunct{\mcitedefaultmidpunct}
{\mcitedefaultendpunct}{\mcitedefaultseppunct}\relax
\EndOfBibitem
\bibitem[Chung \latin{et~al.}(2016)Chung, Chang, Lin, and
  Lin]{chung2016electronic}
Chung,~H.~C.; Chang,~C.~P.; Lin,~C.~Y.; Lin,~M.~F. \emph{Physical Chemistry
  Chemical Physics} \textbf{2016}, \emph{18}, 7573--7616\relax
\mciteBstWouldAddEndPuncttrue
\mciteSetBstMidEndSepPunct{\mcitedefaultmidpunct}
{\mcitedefaultendpunct}{\mcitedefaultseppunct}\relax
\EndOfBibitem
\bibitem[Kim and Kim(2008)Kim, and Kim]{kim2008prediction}
Kim,~W.~Y.; Kim,~K.~S. \emph{Nature nanotechnology} \textbf{2008}, \emph{3},
  408--412\relax
\mciteBstWouldAddEndPuncttrue
\mciteSetBstMidEndSepPunct{\mcitedefaultmidpunct}
{\mcitedefaultendpunct}{\mcitedefaultseppunct}\relax
\EndOfBibitem
\bibitem[Yang \latin{et~al.}(2007)Yang, Cohen, and Louie]{yang2007excitonic}
Yang,~L.; Cohen,~M.~L.; Louie,~S.~G. \emph{Nano letters} \textbf{2007},
  \emph{7}, 3112--3115\relax
\mciteBstWouldAddEndPuncttrue
\mciteSetBstMidEndSepPunct{\mcitedefaultmidpunct}
{\mcitedefaultendpunct}{\mcitedefaultseppunct}\relax
\EndOfBibitem
\bibitem[Lin and Shyu(2000)Lin, and Shyu]{lin2000optical}
Lin,~M.~F.; Shyu,~F.~L. \emph{Journal of the Physical Society of Japan}
  \textbf{2000}, \emph{69}, 3529--3532\relax
\mciteBstWouldAddEndPuncttrue
\mciteSetBstMidEndSepPunct{\mcitedefaultmidpunct}
{\mcitedefaultendpunct}{\mcitedefaultseppunct}\relax
\EndOfBibitem
\bibitem[Li \latin{et~al.}(2008)Li, Qian, Wu, Gu, and Duan]{li2008role}
Li,~Z.; Qian,~H.; Wu,~J.; Gu,~B.~L.; Duan,~W. \emph{Physical review letters}
  \textbf{2008}, \emph{100}, 206802\relax
\mciteBstWouldAddEndPuncttrue
\mciteSetBstMidEndSepPunct{\mcitedefaultmidpunct}
{\mcitedefaultendpunct}{\mcitedefaultseppunct}\relax
\EndOfBibitem
\bibitem[Basu \latin{et~al.}(2008)Basu, Gilbert, Register, Banerjee, and
  MacDonald]{basu2008effect}
Basu,~D.; Gilbert,~M.; Register,~L.; Banerjee,~S.~K.; MacDonald,~A.~H.
  \emph{Applied Physics Letters} \textbf{2008}, \emph{92}, 042114\relax
\mciteBstWouldAddEndPuncttrue
\mciteSetBstMidEndSepPunct{\mcitedefaultmidpunct}
{\mcitedefaultendpunct}{\mcitedefaultseppunct}\relax
\EndOfBibitem
\bibitem[Son \latin{et~al.}(2006)Son, Cohen, and Louie]{son2006energy}
Son,~Y.~W.; Cohen,~M.~L.; Louie,~S.~G. \emph{Physical review letters}
  \textbf{2006}, \emph{97}, 216803\relax
\mciteBstWouldAddEndPuncttrue
\mciteSetBstMidEndSepPunct{\mcitedefaultmidpunct}
{\mcitedefaultendpunct}{\mcitedefaultseppunct}\relax
\EndOfBibitem
\bibitem[Barone \latin{et~al.}(2006)Barone, Hod, and
  Scuseria]{barone2006electronic}
Barone,~V.; Hod,~O.; Scuseria,~G.~E. \emph{Nano letters} \textbf{2006},
  \emph{6}, 2748--2754\relax
\mciteBstWouldAddEndPuncttrue
\mciteSetBstMidEndSepPunct{\mcitedefaultmidpunct}
{\mcitedefaultendpunct}{\mcitedefaultseppunct}\relax
\EndOfBibitem
\bibitem[Ritter and Lyding(2009)Ritter, and Lyding]{ritter2009influence}
Ritter,~K.~A.; Lyding,~J.~W. \emph{Nature materials} \textbf{2009}, \emph{8},
  235--242\relax
\mciteBstWouldAddEndPuncttrue
\mciteSetBstMidEndSepPunct{\mcitedefaultmidpunct}
{\mcitedefaultendpunct}{\mcitedefaultseppunct}\relax
\EndOfBibitem
\bibitem[Chang \latin{et~al.}(2014)Chang, Lin, Lin, Lee, and
  Lin]{chang2014geometric}
Chang,~S.~L.; Lin,~S.~Y.; Lin,~S.~K.; Lee,~C.~H.; Lin,~M.~F. \emph{Scientific
  reports} \textbf{2014}, \emph{4}, 6038\relax
\mciteBstWouldAddEndPuncttrue
\mciteSetBstMidEndSepPunct{\mcitedefaultmidpunct}
{\mcitedefaultendpunct}{\mcitedefaultseppunct}\relax
\EndOfBibitem
\bibitem[Lin \latin{et~al.}(2015)Lin, Chung, Yang, Lin, and Lin]{lin2015adatom}
Lin,~Y.~T.; Chung,~H.~C.; Yang,~P.~H.; Lin,~S.~Y.; Lin,~M.~F. \emph{Physical
  Chemistry Chemical Physics} \textbf{2015}, \emph{17}, 16545--16552\relax
\mciteBstWouldAddEndPuncttrue
\mciteSetBstMidEndSepPunct{\mcitedefaultmidpunct}
{\mcitedefaultendpunct}{\mcitedefaultseppunct}\relax
\EndOfBibitem
\bibitem[Johnson \latin{et~al.}(2010)Johnson, Behnam, Pearton, and
  Ural]{johnson2010hydrogen}
Johnson,~J.~L.; Behnam,~A.; Pearton,~S.; Ural,~A. \emph{Advanced Materials}
  \textbf{2010}, \emph{22}, 4877--4880\relax
\mciteBstWouldAddEndPuncttrue
\mciteSetBstMidEndSepPunct{\mcitedefaultmidpunct}
{\mcitedefaultendpunct}{\mcitedefaultseppunct}\relax
\EndOfBibitem
\bibitem[Lin \latin{et~al.}(2013)Lin, Peng, Xiang, Ruan, Yan, Natelson, and
  Tour]{lin2013graphene}
Lin,~J.; Peng,~Z.; Xiang,~C.; Ruan,~G.; Yan,~Z.; Natelson,~D.; Tour,~J.~M.
  \emph{ACS nano} \textbf{2013}, \emph{7}, 6001--6006\relax
\mciteBstWouldAddEndPuncttrue
\mciteSetBstMidEndSepPunct{\mcitedefaultmidpunct}
{\mcitedefaultendpunct}{\mcitedefaultseppunct}\relax
\EndOfBibitem
\bibitem[Huang \latin{et~al.}(2008)Huang, Chang, and
  Lin]{huang2008magnetoabsorption}
Huang,~Y.~C.; Chang,~C.~P.; Lin,~M.~F. \emph{Physical Review B} \textbf{2008},
  \emph{78}, 115422\relax
\mciteBstWouldAddEndPuncttrue
\mciteSetBstMidEndSepPunct{\mcitedefaultmidpunct}
{\mcitedefaultendpunct}{\mcitedefaultseppunct}\relax
\EndOfBibitem
\bibitem[Sadeghi \latin{et~al.}(2012)Sadeghi, Ahmadi, Mousavi, Ismail, and
  Ghadiry]{sadeghi2012channel}
Sadeghi,~H.; Ahmadi,~M.; Mousavi,~S.; Ismail,~R.; Ghadiry,~M.~H. \emph{Modern
  Physics Letters B} \textbf{2012}, \emph{26}, 1250047\relax
\mciteBstWouldAddEndPuncttrue
\mciteSetBstMidEndSepPunct{\mcitedefaultmidpunct}
{\mcitedefaultendpunct}{\mcitedefaultseppunct}\relax
\EndOfBibitem
\bibitem[Lin \latin{et~al.}(2012)Lin, Chen, Wu, and Lin]{lin2012curvature}
Lin,~C.~Y.; Chen,~S.~C.; Wu,~J.~Y.; Lin,~M.~F. \emph{Journal of the Physical
  Society of Japan} \textbf{2012}, \emph{81}, 064719\relax
\mciteBstWouldAddEndPuncttrue
\mciteSetBstMidEndSepPunct{\mcitedefaultmidpunct}
{\mcitedefaultendpunct}{\mcitedefaultseppunct}\relax
\EndOfBibitem
\bibitem[Chang \latin{et~al.}(2012)Chang, Wu, Yang, and
  Lin]{chang2012curvature}
Chang,~S.~L.; Wu,~B.~R.; Yang,~P.~H.; Lin,~M.~F. \emph{Physical Chemistry
  Chemical Physics} \textbf{2012}, \emph{14}, 16409--16414\relax
\mciteBstWouldAddEndPuncttrue
\mciteSetBstMidEndSepPunct{\mcitedefaultmidpunct}
{\mcitedefaultendpunct}{\mcitedefaultseppunct}\relax
\EndOfBibitem
\bibitem[Chang \latin{et~al.}(2007)Chang, Wu, Chen, and
  Lin]{chang2007deformation}
Chang,~C.~P.; Wu,~B.~R.; Chen,~R.~B.; Lin,~M.~F. \emph{Journal of applied
  physics} \textbf{2007}, \emph{101}, 063506\relax
\mciteBstWouldAddEndPuncttrue
\mciteSetBstMidEndSepPunct{\mcitedefaultmidpunct}
{\mcitedefaultendpunct}{\mcitedefaultseppunct}\relax
\EndOfBibitem
\bibitem[Lin \latin{et~al.}(2015)Lin, Chang, Shyu, Lu, and Lin]{lin2015feature}
Lin,~S.~Y.; Chang,~S.~L.; Shyu,~F.~L.; Lu,~J.~M.; Lin,~M.~F. \emph{Carbon}
  \textbf{2015}, \emph{86}, 207--216\relax
\mciteBstWouldAddEndPuncttrue
\mciteSetBstMidEndSepPunct{\mcitedefaultmidpunct}
{\mcitedefaultendpunct}{\mcitedefaultseppunct}\relax
\EndOfBibitem
\bibitem[Chang \latin{et~al.}(2006)Chang, Huang, Lu, Ho, Li, and
  Lin]{chang2006electronic}
Chang,~C.~P.; Huang,~Y.~C.; Lu,~C.~L.; Ho,~J.~H.; Li,~T.~S.; Lin,~M.~F.
  \emph{Carbon} \textbf{2006}, \emph{44}, 508--515\relax
\mciteBstWouldAddEndPuncttrue
\mciteSetBstMidEndSepPunct{\mcitedefaultmidpunct}
{\mcitedefaultendpunct}{\mcitedefaultseppunct}\relax
\EndOfBibitem
\bibitem[Kan \latin{et~al.}(2007)Kan, Li, Yang, and Hou]{kan2007will}
Kan,~E.~J.; Li,~Z.; Yang,~J.; Hou,~J. \emph{Applied physics letters}
  \textbf{2007}, \emph{91}, 243116\relax
\mciteBstWouldAddEndPuncttrue
\mciteSetBstMidEndSepPunct{\mcitedefaultmidpunct}
{\mcitedefaultendpunct}{\mcitedefaultseppunct}\relax
\EndOfBibitem
\bibitem[Raza and Kan(2008)Raza, and Kan]{raza2008armchair}
Raza,~H.; Kan,~E.~C. \emph{Physical Review B} \textbf{2008}, \emph{77},
  245434\relax
\mciteBstWouldAddEndPuncttrue
\mciteSetBstMidEndSepPunct{\mcitedefaultmidpunct}
{\mcitedefaultendpunct}{\mcitedefaultseppunct}\relax
\EndOfBibitem
\bibitem[Huang \latin{et~al.}(2007)Huang, Chang, and Lin]{huang2007magnetic}
Huang,~Y.~C.; Chang,~C.~P.; Lin,~M.~F. \emph{Nanotechnology} \textbf{2007},
  \emph{18}, 495401\relax
\mciteBstWouldAddEndPuncttrue
\mciteSetBstMidEndSepPunct{\mcitedefaultmidpunct}
{\mcitedefaultendpunct}{\mcitedefaultseppunct}\relax
\EndOfBibitem
\bibitem[Liu \latin{et~al.}(2008)Liu, Wright, Zhang, and Ma]{liu2008strong}
Liu,~J.; Wright,~A.; Zhang,~C.; Ma,~Z. \emph{Applied Physics Letters}
  \textbf{2008}, \emph{93}, 041106\relax
\mciteBstWouldAddEndPuncttrue
\mciteSetBstMidEndSepPunct{\mcitedefaultmidpunct}
{\mcitedefaultendpunct}{\mcitedefaultseppunct}\relax
\EndOfBibitem
\bibitem[Obradovic \latin{et~al.}(2006)Obradovic, Kotlyar, Heinz, Matagne,
  Rakshit, Giles, Stettler, and Nikonov]{obradovic2006analysis}
Obradovic,~B.; Kotlyar,~R.; Heinz,~F.; Matagne,~P.; Rakshit,~T.; Giles,~M.;
  Stettler,~M.; Nikonov,~D. \emph{Applied Physics Letters} \textbf{2006},
  \emph{88}, 142102\relax
\mciteBstWouldAddEndPuncttrue
\mciteSetBstMidEndSepPunct{\mcitedefaultmidpunct}
{\mcitedefaultendpunct}{\mcitedefaultseppunct}\relax
\EndOfBibitem
\bibitem[Saffarzadeh and Farghadan(2011)Saffarzadeh, and
  Farghadan]{saffarzadeh2011spin}
Saffarzadeh,~A.; Farghadan,~R. \emph{Applied Physics Letters} \textbf{2011},
  \emph{98}, 023106\relax
\mciteBstWouldAddEndPuncttrue
\mciteSetBstMidEndSepPunct{\mcitedefaultmidpunct}
{\mcitedefaultendpunct}{\mcitedefaultseppunct}\relax
\EndOfBibitem
\bibitem[Tapaszt{\'o} \latin{et~al.}(2008)Tapaszt{\'o}, Dobrik, Lambin, and
  Bir{\'o}]{tapaszto2008tailoring}
Tapaszt{\'o},~L.; Dobrik,~G.; Lambin,~P.; Bir{\'o},~L.~P. \emph{Nature
  nanotechnology} \textbf{2008}, \emph{3}, 397--401\relax
\mciteBstWouldAddEndPuncttrue
\mciteSetBstMidEndSepPunct{\mcitedefaultmidpunct}
{\mcitedefaultendpunct}{\mcitedefaultseppunct}\relax
\EndOfBibitem
\bibitem[S{\"o}de \latin{et~al.}(2015)S{\"o}de, Talirz, Gr{\"o}ning, Pignedoli,
  Berger, Feng, M{\"u}llen, Fasel, and Ruffieux]{sode2015electronic}
S{\"o}de,~H.; Talirz,~L.; Gr{\"o}ning,~O.; Pignedoli,~C.~A.; Berger,~R.;
  Feng,~X.; M{\"u}llen,~K.; Fasel,~R.; Ruffieux,~P. \emph{Physical Review B}
  \textbf{2015}, \emph{91}, 045429\relax
\mciteBstWouldAddEndPuncttrue
\mciteSetBstMidEndSepPunct{\mcitedefaultmidpunct}
{\mcitedefaultendpunct}{\mcitedefaultseppunct}\relax
\EndOfBibitem
\bibitem[Chen \latin{et~al.}(2013)Chen, De~Oteyza, Pedramrazi, Chen, Fischer,
  and Crommie]{chen2013tuning}
Chen,~Y.~C.; De~Oteyza,~D.~G.; Pedramrazi,~Z.; Chen,~C.; Fischer,~F.~R.;
  Crommie,~M.~F. \emph{ACS nano} \textbf{2013}, \emph{7}, 6123--6128\relax
\mciteBstWouldAddEndPuncttrue
\mciteSetBstMidEndSepPunct{\mcitedefaultmidpunct}
{\mcitedefaultendpunct}{\mcitedefaultseppunct}\relax
\EndOfBibitem
\bibitem[Mao \latin{et~al.}(2013)Mao, Hao, Wei, Yuan, and Zhong]{mao2013edge}
Mao,~Y.; Hao,~W.; Wei,~X.; Yuan,~J.; Zhong,~J. \emph{Applied Surface Science}
  \textbf{2013}, \emph{280}, 698--704\relax
\mciteBstWouldAddEndPuncttrue
\mciteSetBstMidEndSepPunct{\mcitedefaultmidpunct}
{\mcitedefaultendpunct}{\mcitedefaultseppunct}\relax
\EndOfBibitem
\bibitem[Wang \latin{et~al.}(2013)Wang, Xiao, and Li]{wang2013adsorption}
Wang,~Z.; Xiao,~J.; Li,~M. \emph{Applied Physics A} \textbf{2013}, \emph{110},
  235--239\relax
\mciteBstWouldAddEndPuncttrue
\mciteSetBstMidEndSepPunct{\mcitedefaultmidpunct}
{\mcitedefaultendpunct}{\mcitedefaultseppunct}\relax
\EndOfBibitem
\bibitem[da~Rocha \latin{et~al.}(2014)da~Rocha, Clayborne, Koskinen, and
  H{\"a}kkinen]{da2014optical}
da~Rocha,~C.~G.; Clayborne,~P.~A.; Koskinen,~P.; H{\"a}kkinen,~H.
  \emph{Physical Chemistry Chemical Physics} \textbf{2014}, \emph{16},
  3558--3565\relax
\mciteBstWouldAddEndPuncttrue
\mciteSetBstMidEndSepPunct{\mcitedefaultmidpunct}
{\mcitedefaultendpunct}{\mcitedefaultseppunct}\relax
\EndOfBibitem
\bibitem[Uthaisar \latin{et~al.}(2009)Uthaisar, Barone, and
  Peralta]{uthaisar2009lithium}
Uthaisar,~C.; Barone,~V.; Peralta,~J.~E. \emph{Journal of Applied Physics}
  \textbf{2009}, \emph{106}, 113715\relax
\mciteBstWouldAddEndPuncttrue
\mciteSetBstMidEndSepPunct{\mcitedefaultmidpunct}
{\mcitedefaultendpunct}{\mcitedefaultseppunct}\relax
\EndOfBibitem
\bibitem[Kresse and Furthm{\"u}ller(1996)Kresse, and
  Furthm{\"u}ller]{kresse1996efficient}
Kresse,~G.; Furthm{\"u}ller,~J. \emph{Physical Review B} \textbf{1996},
  \emph{54}, 11169\relax
\mciteBstWouldAddEndPuncttrue
\mciteSetBstMidEndSepPunct{\mcitedefaultmidpunct}
{\mcitedefaultendpunct}{\mcitedefaultseppunct}\relax
\EndOfBibitem
\bibitem[Perdew \latin{et~al.}(1996)Perdew, Burke, and
  Ernzerhof]{perdew1996generalized}
Perdew,~J.~P.; Burke,~K.; Ernzerhof,~M. \emph{Physical review letters}
  \textbf{1996}, \emph{77}, 3865\relax
\mciteBstWouldAddEndPuncttrue
\mciteSetBstMidEndSepPunct{\mcitedefaultmidpunct}
{\mcitedefaultendpunct}{\mcitedefaultseppunct}\relax
\EndOfBibitem
\bibitem[Bl{\"o}chl(1994)]{blochl1994projector}
Bl{\"o}chl,~P.~E. \emph{Physical Review B} \textbf{1994}, \emph{50},
  17953\relax
\mciteBstWouldAddEndPuncttrue
\mciteSetBstMidEndSepPunct{\mcitedefaultmidpunct}
{\mcitedefaultendpunct}{\mcitedefaultseppunct}\relax
\EndOfBibitem
\bibitem[Virojanadara \latin{et~al.}(2010)Virojanadara, Watcharinyanon,
  Zakharov, and Johansson]{virojanadara2010epitaxial}
Virojanadara,~C.; Watcharinyanon,~S.; Zakharov,~A.; Johansson,~L.~I.
  \emph{Physical Review B} \textbf{2010}, \emph{82}, 205402\relax
\mciteBstWouldAddEndPuncttrue
\mciteSetBstMidEndSepPunct{\mcitedefaultmidpunct}
{\mcitedefaultendpunct}{\mcitedefaultseppunct}\relax
\EndOfBibitem
\bibitem[Lin \latin{et~al.}(2015)Lin, Gong, Lu, Wu, Wang, Guan, Angell, Chen,
  Yang, and Hwang]{lin2015ultrafast}
Lin,~M.~C.; Gong,~M.; Lu,~B.; Wu,~Y.; Wang,~D.~Y.; Guan,~M.; Angell,~M.;
  Chen,~C.; Yang,~J.; Hwang,~B.~J. \emph{Nature} \textbf{2015}, \emph{520},
  324--328\relax
\mciteBstWouldAddEndPuncttrue
\mciteSetBstMidEndSepPunct{\mcitedefaultmidpunct}
{\mcitedefaultendpunct}{\mcitedefaultseppunct}\relax
\EndOfBibitem
\bibitem[Rani \latin{et~al.}(2013)Rani, Kanakaiah, Dadmal, Rao, and
  Bhavanarushi]{rani2013fluorinated}
Rani,~J.~V.; Kanakaiah,~V.; Dadmal,~T.; Rao,~M.~S.; Bhavanarushi,~S.
  \emph{Journal of The Electrochemical Society} \textbf{2013}, \emph{160},
  A1781--A1784\relax
\mciteBstWouldAddEndPuncttrue
\mciteSetBstMidEndSepPunct{\mcitedefaultmidpunct}
{\mcitedefaultendpunct}{\mcitedefaultseppunct}\relax
\EndOfBibitem
\bibitem[Paek \latin{et~al.}(2008)Paek, Yoo, and Honma]{paek2008enhanced}
Paek,~S.~M.; Yoo,~E.; Honma,~I. \emph{Nano letters} \textbf{2008}, \emph{9},
  72--75\relax
\mciteBstWouldAddEndPuncttrue
\mciteSetBstMidEndSepPunct{\mcitedefaultmidpunct}
{\mcitedefaultendpunct}{\mcitedefaultseppunct}\relax
\EndOfBibitem
\bibitem[Wang \latin{et~al.}(2009)Wang, Shen, Yao, and Park]{wang2009graphene}
Wang,~G.; Shen,~X.; Yao,~J.; Park,~J. \emph{Carbon} \textbf{2009}, \emph{47},
  2049--2053\relax
\mciteBstWouldAddEndPuncttrue
\mciteSetBstMidEndSepPunct{\mcitedefaultmidpunct}
{\mcitedefaultendpunct}{\mcitedefaultseppunct}\relax
\EndOfBibitem
\bibitem[Sugawara \latin{et~al.}(2006)Sugawara, Sato, Souma, Takahashi, and
  Suematsu]{sugawara2006fermi}
Sugawara,~K.; Sato,~T.; Souma,~S.; Takahashi,~T.; Suematsu,~H. \emph{Physical
  Review B} \textbf{2006}, \emph{73}, 045124\relax
\mciteBstWouldAddEndPuncttrue
\mciteSetBstMidEndSepPunct{\mcitedefaultmidpunct}
{\mcitedefaultendpunct}{\mcitedefaultseppunct}\relax
\EndOfBibitem
\bibitem[Gr{\"u}neis \latin{et~al.}(2008)Gr{\"u}neis, Attaccalite, Pichler,
  Zabolotnyy, Shiozawa, Molodtsov, Inosov, Koitzsch, Knupfer, and
  Schiessling]{gruneis2008electron}
Gr{\"u}neis,~A.; Attaccalite,~C.; Pichler,~T.; Zabolotnyy,~V.; Shiozawa,~H.;
  Molodtsov,~S.; Inosov,~D.; Koitzsch,~A.; Knupfer,~M.; Schiessling,~J.
  \emph{Physical review letters} \textbf{2008}, \emph{100}, 037601\relax
\mciteBstWouldAddEndPuncttrue
\mciteSetBstMidEndSepPunct{\mcitedefaultmidpunct}
{\mcitedefaultendpunct}{\mcitedefaultseppunct}\relax
\EndOfBibitem
\bibitem[Ohta \latin{et~al.}(2007)Ohta, Bostwick, McChesney, Seyller, Horn, and
  Rotenberg]{ohta2007interlayer}
Ohta,~T.; Bostwick,~A.; McChesney,~J.~L.; Seyller,~T.; Horn,~K.; Rotenberg,~E.
  \emph{Physical Review Letters} \textbf{2007}, \emph{98}, 206802\relax
\mciteBstWouldAddEndPuncttrue
\mciteSetBstMidEndSepPunct{\mcitedefaultmidpunct}
{\mcitedefaultendpunct}{\mcitedefaultseppunct}\relax
\EndOfBibitem
\bibitem[Siegel \latin{et~al.}(2013)Siegel, Regan, Fedorov, Zettl, and
  Lanzara]{siegel2013charge}
Siegel,~D.~A.; Regan,~W.; Fedorov,~A.~V.; Zettl,~A.; Lanzara,~A. \emph{Physical
  review letters} \textbf{2013}, \emph{110}, 146802\relax
\mciteBstWouldAddEndPuncttrue
\mciteSetBstMidEndSepPunct{\mcitedefaultmidpunct}
{\mcitedefaultendpunct}{\mcitedefaultseppunct}\relax
\EndOfBibitem
\bibitem[Bostwick \latin{et~al.}(2007)Bostwick, Ohta, Seyller, Horn, and
  Rotenberg]{bostwick2007quasiparticle}
Bostwick,~A.; Ohta,~T.; Seyller,~T.; Horn,~K.; Rotenberg,~E. \emph{Nature
  Physics} \textbf{2007}, \emph{3}, 36--40\relax
\mciteBstWouldAddEndPuncttrue
\mciteSetBstMidEndSepPunct{\mcitedefaultmidpunct}
{\mcitedefaultendpunct}{\mcitedefaultseppunct}\relax
\EndOfBibitem
\bibitem[Ohta \latin{et~al.}(2006)Ohta, Bostwick, Seyller, Horn, and
  Rotenberg]{ohta2006controlling}
Ohta,~T.; Bostwick,~A.; Seyller,~T.; Horn,~K.; Rotenberg,~E. \emph{Science}
  \textbf{2006}, \emph{313}, 951--954\relax
\mciteBstWouldAddEndPuncttrue
\mciteSetBstMidEndSepPunct{\mcitedefaultmidpunct}
{\mcitedefaultendpunct}{\mcitedefaultseppunct}\relax
\EndOfBibitem
\bibitem[Coletti \latin{et~al.}(2013)Coletti, Forti, Principi, Emtsev,
  Zakharov, Daniels, Daas, Chandrashekhar, Ouisse, and
  Chaussende]{coletti2013revealing}
Coletti,~C.; Forti,~S.; Principi,~A.; Emtsev,~K.~V.; Zakharov,~A.~A.;
  Daniels,~K.~M.; Daas,~B.~K.; Chandrashekhar,~M.; Ouisse,~T.; Chaussende,~D.
  \emph{Physical Review B} \textbf{2013}, \emph{88}, 155439\relax
\mciteBstWouldAddEndPuncttrue
\mciteSetBstMidEndSepPunct{\mcitedefaultmidpunct}
{\mcitedefaultendpunct}{\mcitedefaultseppunct}\relax
\EndOfBibitem
\bibitem[Zhou \latin{et~al.}(2008)Zhou, Siegel, Fedorov, and
  Lanzara]{zhou2008metal}
Zhou,~S.; Siegel,~D.; Fedorov,~A.; Lanzara,~A. \emph{Physical review letters}
  \textbf{2008}, \emph{101}, 086402\relax
\mciteBstWouldAddEndPuncttrue
\mciteSetBstMidEndSepPunct{\mcitedefaultmidpunct}
{\mcitedefaultendpunct}{\mcitedefaultseppunct}\relax
\EndOfBibitem
\bibitem[Schulte \latin{et~al.}(2013)Schulte, Vinogradov, Ng, M{\aa}rtensson,
  and Preobrajenski]{schulte2013bandgap}
Schulte,~K.; Vinogradov,~N.; Ng,~M.~L.; M{\aa}rtensson,~N.; Preobrajenski,~A.
  \emph{Applied Surface Science} \textbf{2013}, \emph{267}, 74--76\relax
\mciteBstWouldAddEndPuncttrue
\mciteSetBstMidEndSepPunct{\mcitedefaultmidpunct}
{\mcitedefaultendpunct}{\mcitedefaultseppunct}\relax
\EndOfBibitem
\bibitem[Klusek(1999)]{klusek1999investigations}
Klusek,~Z. \emph{Applied surface science} \textbf{1999}, \emph{151},
  251--261\relax
\mciteBstWouldAddEndPuncttrue
\mciteSetBstMidEndSepPunct{\mcitedefaultmidpunct}
{\mcitedefaultendpunct}{\mcitedefaultseppunct}\relax
\EndOfBibitem
\bibitem[Li \latin{et~al.}(2009)Li, Luican, and Andrei]{li2009scanning}
Li,~G.; Luican,~A.; Andrei,~E.~Y. \emph{Physical Review Letters} \textbf{2009},
  \emph{102}, 176804\relax
\mciteBstWouldAddEndPuncttrue
\mciteSetBstMidEndSepPunct{\mcitedefaultmidpunct}
{\mcitedefaultendpunct}{\mcitedefaultseppunct}\relax
\EndOfBibitem
\bibitem[Luican \latin{et~al.}(2011)Luican, Li, Reina, Kong, Nair, Novoselov,
  Geim, and Andrei]{luican2011single}
Luican,~A.; Li,~G.; Reina,~A.; Kong,~J.; Nair,~R.; Novoselov,~K.~S.;
  Geim,~A.~K.; Andrei,~E. \emph{Physical review letters} \textbf{2011},
  \emph{106}, 126802\relax
\mciteBstWouldAddEndPuncttrue
\mciteSetBstMidEndSepPunct{\mcitedefaultmidpunct}
{\mcitedefaultendpunct}{\mcitedefaultseppunct}\relax
\EndOfBibitem
\bibitem[Li \latin{et~al.}(2010)Li, Luican, Dos~Santos, Neto, Reina, Kong, and
  Andrei]{li2010observation}
Li,~G.; Luican,~A.; Dos~Santos,~J.~L.; Neto,~A.~C.; Reina,~A.; Kong,~J.;
  Andrei,~E. \emph{Nature Physics} \textbf{2010}, \emph{6}, 109--113\relax
\mciteBstWouldAddEndPuncttrue
\mciteSetBstMidEndSepPunct{\mcitedefaultmidpunct}
{\mcitedefaultendpunct}{\mcitedefaultseppunct}\relax
\EndOfBibitem
\bibitem[Cherkez \latin{et~al.}(2015)Cherkez, de~Laissardiere, Mallet, and
  Veuillen]{cherkez2015van}
Cherkez,~V.; de~Laissardiere,~G.~T.; Mallet,~P.; Veuillen,~J.~Y. \emph{Physical
  Review B} \textbf{2015}, \emph{91}, 155428\relax
\mciteBstWouldAddEndPuncttrue
\mciteSetBstMidEndSepPunct{\mcitedefaultmidpunct}
{\mcitedefaultendpunct}{\mcitedefaultseppunct}\relax
\EndOfBibitem
\bibitem[Wilder \latin{et~al.}(1998)Wilder, Venema, Rinzler, Smalley, and
  Dekker]{wilder1998electronic}
Wilder,~J.~W.; Venema,~L.~C.; Rinzler,~A.~G.; Smalley,~R.~E.; Dekker,~C.
  \emph{Nature} \textbf{1998}, \emph{391}, 59--62\relax
\mciteBstWouldAddEndPuncttrue
\mciteSetBstMidEndSepPunct{\mcitedefaultmidpunct}
{\mcitedefaultendpunct}{\mcitedefaultseppunct}\relax
\EndOfBibitem
\bibitem[Odom \latin{et~al.}(1998)Odom, Huang, Kim, and Lieber]{odom1998atomic}
Odom,~T.~W.; Huang,~J.~L.; Kim,~P.; Lieber,~C.~M. \emph{Nature} \textbf{1998},
  \emph{391}, 62--64\relax
\mciteBstWouldAddEndPuncttrue
\mciteSetBstMidEndSepPunct{\mcitedefaultmidpunct}
{\mcitedefaultendpunct}{\mcitedefaultseppunct}\relax
\EndOfBibitem
\end{mcitethebibliography}

\end{document}